\documentclass[12pt]{article}
\usepackage{amsmath}
\usepackage{amssymb}
\usepackage{amsthm}
\usepackage{graphicx}
\usepackage{lineno}
\usepackage{float}
\usepackage[all]{xy}
\usepackage{mathrsfs}

\usepackage{xcolor}
\usepackage{tikz}
\usetikzlibrary{tqft}

\newtheorem*{theorem*}{Theorem}

\newtheorem{example}{Example}

\renewcommand{\H}{\mathcal H}

\newcommand{\ovl}{\overline}

\newcommand{\wti}{\widetilde}

\begin{document}

\title{\bf 
Communication protocols and QECC from the perspective of TQFT, Part II: \\
QECCs as spacetimes}

\author{{Chris Fields$^a$, James F. Glazebrook$^{b,c}$ and Antonino Marcian\`{o}$^{d,e,f}$}\\ \\
{\it$^a$ Allen Discovery Center, Tufts University, Medford, MA 02155 USA}\\
{fieldsres@gmail.com}\\
{ORCID: 0000-0002-4812-0744}\\
{\it$^b$ Department of Mathematics and Computer Science,} \\
{\it Eastern Illinois University, Charleston, IL 61920 USA} \\
{\it$^c$ Adjunct Faculty, Department of Mathematics,}\\
{\it University of Illinois at Urbana-Champaign, Urbana, IL 61801 USA}\\
{jfglazebrook@eiu.edu}\\
{ORCID: 0000-0001-8335-221X}\\
{\it$^d$ Center for Field Theory and Particle Physics \& Department of Physics} \\
{\it Fudan University, Shanghai, CHINA} \\
{marciano@fudan.edu.cn} \\
{\it$^e$ Laboratori Nazionali di Frascati INFN, Frascati (Rome), Italy, EU} \\
{marciano@lnf.infn.it} \\
{\it$^f$ INFN sezione Roma ``Tor Vergata'', 00133 Rome, Italy, EU} \\
{ORCID: 0000-0003-4719-110X}
}

\maketitle

{\bf Abstract:} \\

Topological quantum field theories (TQFTs) provide a general, minimal-assumption language for describing quantum-state preparation and measurement. They therefore provide a general language in which to express multi-agent communication protocols, e.g. local operations, classical communication (LOCC) protocols. In the accompanying Part I, we construct LOCC protocols using TQFT, and show that LOCC protocols induce quantum error-correcting codes (QECCs) on the agent-environment boundary.  Such QECCs can be regarded as implementing or inducing the emergence of spacetimes on such boundaries. Here we investigate this connection between inter-agent communication and spacetime, exploiting different realizations of TQFT. We delve into TQFTs that support on their boundaries spin-networks as computational systems: these are known as topological quantum neural networks (TQNNs). TQNNs, which have a natural representation as tensor networks, implement QECC. We recognize into the HaPPY code a paradigmatic example. We then show how generic QECCs, as bulk- boundary codes, induce effective spacetimes. The effective spatial and temporal separations that take place in QECC enables LOCC protocols between spatially separated observers. We then consider the implementation of QECCs in BF and Chern-Simons theories, and show that QECC-induced spacetimes provide the classical redundancy required for LOCC. Finally, we consider topological M-theory as an implementation of QECC in higher spacetime dimensions.

\tableofcontents

\section{Introduction}

We have previously \cite{fgm:22} shown that sequential observations of any finite physical system $S$ that employ either one or some sequence of quantum reference frames (QRFs) \cite{aharonov:84, bartlett:07} induce a topological quantum field theory (TQFT) \cite{atiyah:88, quinn:95} on $S$. A TQFT being the ``default'' physical theory induced on any system by measurement is not surprising: it reflects the fact that spacetime coordinates must, in any sequence of measurements, be specified by particular QRFs.  An immediate consequence of this construction is that any effective field theory (EFT) defined on $S$ must be gauge-invariant \cite{fgm:22, addazi:21}.

In the accompanying Part I of this paper, we showed how any sequence of actions of, or equivalently, operations with, one or more QRFs can simply be identified with the TQFT that it induces.  That this should be the case is also not surprising.  A QRF is a physical system $Q$ with some internal dynamics representable by a Hamiltonian $H_Q$.  Hence treating it as isolated, it evolves in time $t$ via a unitary operator $\mathcal{P}_Q = \exp[(\imath / \hbar) H_Q t]$.  The ``boundaries'' between which $Q$ can be considered isolated are precisely its actions on $S$ at some sequential times $t_i, t_j, \dots t_k$.  Evolving $S$ through time and evolving $Q$ through time are, therefore, operationally the same process: they yield precisely the same data, the data obtained by acting with $Q$ on $S$ at $t_i, t_j, \dots t_k$.  This is in fact a bulk-boundary duality, with $Q$ the bulk and $\mathcal{H}_S$ the boundary \cite{fgm:22, fgm:22a}.  

Using this identification, we then constructed a purely topological representation of multi-party communication protocols involving both quantum and classical resources, i.e. Local Operations, Classical Communication (LOCC) protocols \cite{chitambar:14}, and demonstrated that quantum Darwinism \cite{zurek:06, zurek:09}, originally formulated to explain the emergence of a ``public'' quantum-to-classical transition, simply describes a LOCC protocol.  Using this representation of LOCC together with generic requirements \cite{knill:97} for a quantum error-correcting code (QECC), we then established the main result of Part I: that interactions meeting the requirements to implement a LOCC protocol generically induce QECCs.  In practice, the converse also holds: using a QECC requires agreement about choice of basis, and hence requires a LOCC protocol.  Employing quantum Darwinism and Bell/EPR experiments \cite{aspect:81, georgescu:21} as examples, we investigated how QECCs effectively convert quantum entanglement into classical redundancy, noting that in both cases the redundant classical information, in the form of reports of observational outcomes, is encoded at multiple, distinct spatial locations.  This is consistent with Einstein's \cite{einstein:48} characterization of spacetime as a resource for separability, and hence a resource for classical redundancy.

In this Part II, we further explore the quantum origins of this relation between spatial separation and classical redundancy, and show how QECCs supporting LOCC protocols generically induce spacetimes on the boundaries between interacting quantum systems.  To make this result more intuitive, consider the process of assigning spatial coordinates to an ordinary, macroscopic object.  The spatial coordinates are not intrinsic to the object, but rather are ancillary: we can assign whatever coordinates to the object we wish.  The symmetries of space formally encode the fact that assigning these coordinates in different ways has no effect on the definition of the object, in particular, no effect on its Hilbert space or its internal Hamiltonian.  It remains the ``same thing'' and exhibits the ``same behavior'' when given a different spatial label.  The spatial labels thus provide a source of redundancy: we can consider a set of ``slices'' distinguished by a time label in which the same object appears with distinct labels indicating distinct spatial locations.  This is a redundant (in time) representation of the object's Hilbert space and internal Hamiltonian, which have not changed.  By replacing the Hilbert space and internal Hamiltonian with operators $\uparrow$ and $\downarrow$ that respectively ``create'' or ``destroy'' them, and assigning the spatial coordinates to these operator pairs with the requirement that integrating the operator pairs over the coordinates always yields one object, we have converted the object into a quantum field.  Note what this involves: we must drop the classical assumption that the object persists through arbitrarily fine-grained time with an assumption of discreteness.  It is a natural extension of this operator-field representation to allow the spatial integral to represent $N > 1$ ``copies'' of the object.  The spatial coordinates here give us the extra degree of freedom needed to escape the no-cloning theorem, i.e. we can represent $N$ field excitations with some state component, e.g. $z$-spin, that remains unspecified locally but constrained globally.  Hence we see clearly the role of space in a QFT: spatial separation renders the ``copies'' mutually separable, and hence enables classical redundancy \cite{addazi:21}.  It is natural, therefore, to think of space as a macroscopic, observationally-accessible consequence of an underlying source of redundancy, i.e. a reservoir of entanglement that implements a QECC.

This intuitive connection between QECCs and spacetime has been formalized in various ways; several approaches formulated in the AdS/CFT setting are reviewed in \cite{bain:20}.  The methods of \cite{almheiri:15,pastawski:15}, for example, involve spatial coordinates in the bulk emerging from the functional requirements of implementing a QECC, while \cite{vanraamsdonk:11} suggests that the connectivity of spacetime in the bulk is related to the entanglement structure of the codespace $\H_C$ associated with a boundary CFT. As pointed out in \cite{bain:20}, the HaPPY code of \cite{pastawski:15} could be interpreted as a fundamental discrete spin system with both the boundary and the bulk emerging in a continuum limit.

Here we first briefly review the representation of LOCC protocols as TQFTs developed in Part I, and note that these TQFTs provide a general representation of the QECCs induced by LOCC protocols.  Such LOCC-induced QECCs are completely generic, and are all bulk-boundary codes; we discuss the QECC developed in \cite{almheiri:15} as an example.  We then turn to the formal setting of TQFTs in which the boundaries are represented by spin networks, which when interpreted as computational systems can be characterized as topological quantum neural networks (TQNNs, \cite{fgm:22, mar:20, mar:22}).  Such TQNNs have a natural representation as tensor networks, which can, in turn, be seen as implementing QECCs; we discuss the HaPPY code as an example.  We show how these generic QECCs induce effective spacetimes; in particular, how they induce effective spatial and temporal separations between otherwise-indistinguishable observational outcomes, thus enabling LOCC between spatially separated observers.  We consider the implementation of QECCs in BF and Chern-Simons (CS) theories, showing how the QECC-induced spacetimes provide the classical redundancy required for LOCC.  Noting that fault-tolerant quantum computation requires a QECC \cite{nc:00}, we discuss the representation of quantum algorithms as TQNNs. Finally, we show in \S \ref{m-theory} how these same considerations can be reformulated in the language of topological M-theory, and provide several specific examples.  We conclude that representing multi-party communication using TQFTs -- and therefore TQNNs -- provides a new perspective on the connection between redundancy and spacetime, and suggests, consistent with remarks made already in \cite{addazi:21}, that redundancy and spacetime may at some fundamental level be the same concept.

\section{From TQFTs to QECCs to spacetimes: generic approach and three dimensional implementations} \label{qecc}

\subsection{LOCC protocols and their associated QECCs} \label{qecc-intro}

Recall from Part I that a LOCC protocol is an observational setting in which two or more observers interact with a quantum state while coordinating and/or reporting their activities via a classical communication channel.  We can represent a generic, two-observer LOCC protocol by Diagram \eqref{locc-diag} (Part I, Diagram 6):
\begin{equation} \label{locc-diag}
\begin{gathered}
\begin{tikzpicture}[every tqft/.append style={transform shape}]
\draw[rotate=90] (0,0) ellipse (2.8cm and 1 cm);
\node[above] at (0,1.9) {$\mathscr{B}$};
\draw [thick] (-0.2,1.9) arc [radius=1, start angle=90, end angle= 270];
\draw [thick] (-0.2,1.3) arc [radius=0.4, start angle=90, end angle= 270];
\draw[rotate=90,fill=green,fill opacity=1] (1.6,0.2) ellipse (0.3 cm and 0.2 cm);
\draw[rotate=90,fill=green,fill opacity=1] (0.2,0.2) ellipse (0.3 cm and 0.2 cm);
\draw [thick] (-0.2,-0.3) arc [radius=1, start angle=90, end angle= 270];
\draw [thick] (-0.2,-0.9) arc [radius=0.4, start angle=90, end angle= 270];
\draw[rotate=90,fill=green,fill opacity=1] (-0.6,0.2) ellipse (0.3 cm and 0.2 cm);
\draw[rotate=90,fill=green,fill opacity=1] (-2.0,0.2) ellipse (0.3 cm and 0.2 cm);
\draw [rotate=180, thick, dashed] (-0.2,0.9) arc [radius=0.7, start angle=90, end angle= 270];
\draw [rotate=180, thick, dashed] (-0.2,0.3) arc [radius=0.1, start angle=90, end angle= 270];
\draw [thick] (-0.2,0.5) -- (0,0.5);
\draw [thick] (-0.2,-0.1) -- (0,-0.1);
\draw [thick] (-0.2,-0.9) -- (0,-0.9);
\draw [thick] (-0.2,-0.3) -- (0,-0.3);
\draw [thick, dashed] (0,0.5) -- (0.2,0.5);
\draw [thick, dashed] (0,-0.1) -- (0.2,-0.1);
\draw [thick, dashed] (0,-0.9) -- (0.2,-0.9);
\draw [thick, dashed] (0,-0.3) -- (0.2,-0.3);
\node[above] at (-3,1.7) {Alice};
\node[above] at (2.8,1.7) {Bob};
\draw [ultra thick, white] (-0.9,1.5) -- (-0.7,1.5);
\draw [ultra thick, white] (-1,1.3) -- (-0.8,1.3);
\draw [ultra thick, white] (-1,1.1) -- (-0.8,1.1);
\draw [ultra thick, white] (-1,0.9) -- (-0.8,0.9);
\draw [ultra thick, white] (-1.1,0.7) -- (-0.8,0.7);
\draw [ultra thick, white] (-1.1,0.5) -- (-0.8,0.5);
\draw [ultra thick, white] (-1,-0.9) -- (-0.8,-0.9);
\draw [ultra thick, white] (-1,-1.1) -- (-0.8,-1.1);
\draw [ultra thick, white] (-1,-1.3) -- (-0.8,-1.3);
\draw [ultra thick, white] (-0.9,-1.5) -- (-0.7,-1.5);
\draw [ultra thick, white] (-0.9,-1.7) -- (-0.7,-1.7);
\draw [ultra thick, white] (-0.8,-1.9) -- (-0.6,-1.9);
\draw [ultra thick, white] (-0.8,-2.1) -- (-0.6,-2.1);
\node[above] at (-1.3,1.4) {$Q_1$};
\node[above] at (-1.3,-2.4) {$Q_2$};
\draw [rotate=180, thick] (-0.2,2.3) arc [radius=2.1, start angle=90, end angle= 270];
\draw [rotate=180, thick] (-0.2,1.7) arc [radius=1.5, start angle=90, end angle= 270];
\draw [thick] (-0.2,1.9) -- (0,1.9);
\draw [thick] (-0.2,1.3) -- (0,1.3);
\draw [thick, dashed] (0.2,1.9) -- (0,1.9);
\draw [thick, dashed] (0.2,1.3) -- (0,1.3);
\draw [thick] (-0.2,-1.7) -- (0,-1.7);
\draw [thick] (-0.2,-2.3) -- (0,-2.3);
\draw [thick, dashed] (0.2,-1.7) -- (0,-1.7);
\draw [thick, dashed] (0.2,-2.3) -- (0,-2.3);
\draw [ultra thick, white] (0.3,2) -- (0.3,1.2);
\draw [ultra thick, white] (0.5,2) -- (0.5,1.2);
\draw [ultra thick, white] (0.7,1.9) -- (0.7,1.1);
\draw [ultra thick, white] (0.3,-2.4) -- (0.3,-1.5);
\draw [ultra thick, white] (0.5,-2.4) -- (0.5,-1.5);
\draw [ultra thick, white] (0.7,-1.8) -- (0.7,-1.5);
\node[above] at (4.5,-2.4) {Classical channel};
\draw [thick, ->] (2.9,-2) -- (0.7,-0.8);
\node[above] at (4.5,-1.4) {Quantum channel};
\draw [thick, ->] (2.9,-0.9) -- (2.3,-0.6);
\end{tikzpicture}
\end{gathered}
\end{equation}
\noindent
where $A_1$ and $A_2$ are the observers, $Q_1$ and $Q_2$ are their respective composite read/write QRFs, and $B$ (``Bob'') is the ``environment'' -- the complement of $A = A_1 A_2$ -- that implements both quantum and classical channels.  Both quantum and classical channels are physically implemented by quantum processes that, using the methods of \cite{fgm:22}, can be represented as TQFTs; the classical channel is distinguished by the requirements that 1) $A_1$ and $A_2$ read from and write to it alternately, and 2) writes to this channel are thermodynamically irreversible.

As shown in Part I, quantum Darwinism \cite{zurek:06, zurek:09} provides a generic model of a LOCC protocol in which two or more observers jointly measure the state $|X \rangle$ of some quantum system $X$.  They employ their classical channel to agree to deploy QRFs that identify and measure $X$ -- as opposed to some other system -- and to compare their results.  We can represent the quantum channel between observers $i$ and $j$ as $E_{X_i} X E_{X_j}$, where $E_{X_i}$ and $E_{X_j}$ are non-overlapping components of the ``environment'' $E_X$ of $X$, i.e. of $B \setminus X$ in Diagram \eqref{locc-diag}.  To assure separability of the observers and hence classical redundancy of their results, the interactions between the observers and their respective environmental fragments must be mutually non-disturbing.  As also shown in Part I, a Bell/EPR experiment \cite{aspect:81, georgescu:21} can be considered an example of two-observer quantum Darwinism in which each observer can manipulate one component of a two-component state.  Here manipulations of the state by one observer affect, via the quantum channel, the observations recorded by the other observer.

Using the methods of \cite{knill:97}, we then showed in Part I that the quantum channel in \eqref{locc-diag} implements a QECC, with the codespace $\mathcal{C}$ defined as in Diagram \eqref{locc-qecc} (Part I, Diagram 18):
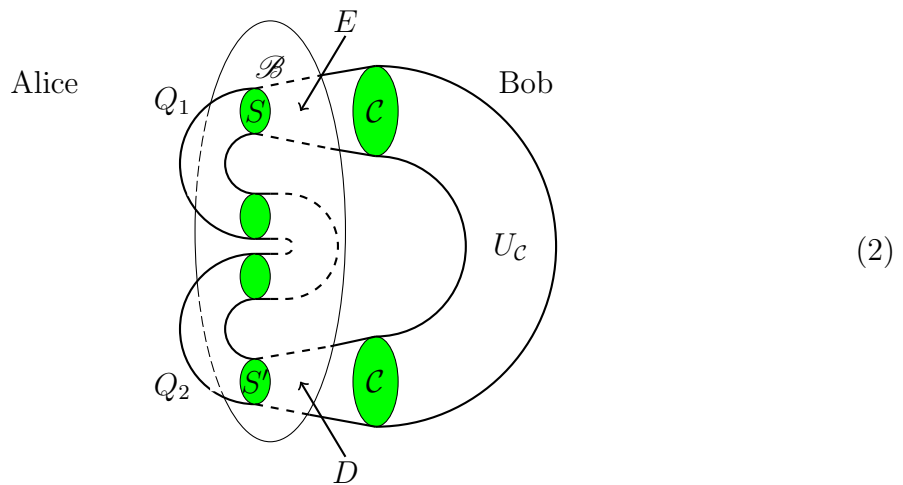
\begin{equation} \label{locc-qecc}
\begin{gathered}
\begin{tikzpicture}[every tqft/.append style={transform shape}]
\draw[rotate=90] (0,0) ellipse (2.8cm and 1 cm);
\node[above] at (0,1.9) {$\mathscr{B}$};
\draw [thick] (-0.2,1.9) arc [radius=1, start angle=90, end angle= 270];
\draw [thick] (-0.2,1.3) arc [radius=0.4, start angle=90, end angle= 270];
\draw[rotate=90,fill=green,fill opacity=1] (1.6,0.2) ellipse (0.3 cm and 0.2 cm);
\draw[rotate=90,fill=green,fill opacity=1] (0.2,0.2) ellipse (0.3 cm and 0.2 cm);
\draw [thick] (-0.2,-0.3) arc [radius=1, start angle=90, end angle= 270];
\draw [thick] (-0.2,-0.9) arc [radius=0.4, start angle=90, end angle= 270];
\draw[rotate=90,fill=green,fill opacity=1] (-0.6,0.2) ellipse (0.3 cm and 0.2 cm);
\draw[rotate=90,fill=green,fill opacity=1] (-2.0,0.2) ellipse (0.3 cm and 0.2 cm);
\draw [rotate=180, thick, dashed] (-0.2,0.9) arc [radius=0.7, start angle=90, end angle= 270];
\draw [rotate=180, thick, dashed] (-0.2,0.3) arc [radius=0.1, start angle=90, end angle= 270];
\draw [thick] (-0.2,0.5) -- (0,0.5);
\draw [thick] (-0.2,-0.1) -- (0,-0.1);
\draw [thick] (-0.2,-0.9) -- (0,-0.9);
\draw [thick] (-0.2,-0.3) -- (0,-0.3);
\draw [thick, dashed] (0,0.5) -- (0.2,0.5);
\draw [thick, dashed] (0,-0.1) -- (0.2,-0.1);
\draw [thick, dashed] (0,-0.9) -- (0.2,-0.9);
\draw [thick, dashed] (0,-0.3) -- (0.2,-0.3);
\node[above] at (-3,1.7) {Alice};
\node[above] at (3.4,1.7) {Bob};
\draw [ultra thick, white] (-0.9,1.5) -- (-0.7,1.5);
\draw [ultra thick, white] (-1,1.3) -- (-0.8,1.3);
\draw [ultra thick, white] (-1,1.1) -- (-0.8,1.1);
\draw [ultra thick, white] (-1,0.9) -- (-0.8,0.9);
\draw [ultra thick, white] (-1.1,0.7) -- (-0.8,0.7);
\draw [ultra thick, white] (-1.1,0.5) -- (-0.8,0.5);
\draw [ultra thick, white] (-1,-0.9) -- (-0.8,-0.9);
\draw [ultra thick, white] (-1,-1.1) -- (-0.8,-1.1);
\draw [ultra thick, white] (-1,-1.3) -- (-0.8,-1.3);
\draw [ultra thick, white] (-0.9,-1.5) -- (-0.7,-1.5);
\draw [ultra thick, white] (-0.9,-1.7) -- (-0.7,-1.7);
\draw [ultra thick, white] (-0.8,-1.9) -- (-0.6,-1.9);
\draw [ultra thick, white] (-0.8,-2.1) -- (-0.6,-2.1);
\node[above] at (-1.3,1.4) {$Q_1$};
\node[above] at (-1.3,-2.4) {$Q_2$};
\draw[rotate=90,fill=green,fill opacity=1] (1.6,-1.4) ellipse (0.6 cm and 0.3 cm);
\draw[rotate=90,fill=green,fill opacity=1] (-2.0,-1.4) ellipse (0.6 cm and 0.3 cm);
\draw [rotate=180, thick] (-1.4,2.6) arc [radius=2.4, start angle=90, end angle= 270];
\draw [rotate=180, thick] (-1.4,1.4) arc [radius=1.2, start angle=90, end angle= 270];
\draw [thick, dashed] (-0.2,1.9) -- (0.7,2.08);
\draw [thick] (0.7,2.08) -- (1.4,2.2);
\draw [thick, dashed] (-0.2,1.3) -- (0.8,1.1);
\draw [thick] (0.8,1.1) -- (1.4,1);
\draw [thick, dashed] (-0.2,-1.7) -- (0.8,-1.53);
\draw [thick] (0.8,-1.53) -- (1.4,-1.4);
\draw [thick, dashed] (-0.2,-2.3) -- (0.52,-2.44);
\draw [thick] (0.52,-2.44) -- (1.4,-2.6);
\node at (-0.2,1.6) {$S$};
\node at (-0.2,-2) {$S^{\prime}$};
\node at (1.4,1.6) {$\mathcal{C}$};
\node at (1.4,-2) {$\mathcal{C}$};
\node at (3.2,-0.2) {$U_{\mathcal{C}}$};
\node at (1,2.8) {$E$};
\draw [thick, ->] (1,2.6) -- (0.4,1.6);
\node at (1,-3.2) {$D$};
\draw [thick, ->] (1,-3) -- (0.4,-2);
\end{tikzpicture}
\end{gathered}
\end{equation}
\noindent
Both the sectors $S$ and $S^{\prime}$ of the $A$-$B$ boundary $\mathscr{B}$ and the codespace $\mathcal{C}$ can, without loss of generality, be represented as finite qubit arrays \cite{fgm:22, fgm:22a}.  As shown in Part I, the sectors $S$ and $S^{\prime}$ must be separable if they are to correspond to observers able to make independent observations.  The quantum channel $E U_{\mathcal{C}} D$ increases (via the encoding $E$) and then decreases (via the decoding $D$) the dimensionality with which the state $|q_i \rangle$ of any qubit $q_i$ in the boundary sector $S$ is represented by qubits of the bulk, effectively entangling $q_i$ with some qubit $q^{\prime}_i$ in $S^{\prime}$.  Classical data encoded by $A_1$ on $S_1$ can, therefore, be recovered, up to perturbations imposed by $B$, by $A_2$ from $S_2$.  This classical redundancy between sectors is enabled by the requirement that $A_1$ and $A_2$ be mutually separable, as discussed in Part I.  In practical settings -- e.g. in a Bell/EPR experiment -- the encoding of classical data may be implemented by instrument settings that are, effectively, communicated to the other observer as observable perturbations of the shared quantum state $|X \rangle$.

Clearly nothing prevents the extension of quantum Darwinism to arbitrary, finite numbers of observers; indeed this is contemplated already in \cite{zurek:06, zurek:09}.  Hence we can regard the boundary $\mathscr{B}$ as a manifold with some arbitrary, finite tessellation, and regard each tile as a sector with an associated observer.  In this case, separability of $\mathscr{B}$ as a manifold corresponds to separability of the associated observers, and allows the observers to each, independently, encode or report classical data.  The bulk $B$ supports error correction for these classical data -- and hence enables classical redundancy -- if it implements a codespace $\mathcal{C}$ that the $i^{th}$ observer can access via encoding and decoding operators $E_i$ and $D_i$ as depicted in Diagram \eqref{locc-qecc}.  Hence the possibility of redundant encoding on all of $\mathscr{B}$ depends on $B$ implementing a single QECC equally accessible from every boundary sector.  Bulk-boundary codes, e.g. as induced by AdS/CFT, implement precisely this condition.

\begin{example}{\bf Bulk-boundary codes, entanglement wedges, and scrambling}

{\rm Diagram \eqref{locc-qecc} raises two obvious questions: 1) how large does $\mathcal{C}$ have to be to protect data encoded on $S$, and 2) what is the relationship between the time evolution $U_{\mathcal{C}}$ of the codespace and the time evolution of the bulk $B$ that implements it?  The second of these questions can be expressed, in the language of \cite{knill:97}, as the question of what perturbations $B_{\alpha}$ we can expect $B$ to impose on $\mathcal{C}$.  Almheiri, Dong and Harlow \cite{almheiri:15} address these questions in the context of AdS/CFT, and with the assumption that the $B_{\alpha}$ are strong enough to erase some of the information in $\mathcal{C}$. i.e. that $B_{\alpha} : |\uparrow \rangle, |\downarrow \rangle \mapsto |0 \rangle =_{Def} (|\uparrow \rangle \pm |\downarrow \rangle)/ \surd 2$ for some subset $e$ of the qubits in $\mathcal{C}$.  They then ask how large $\mathcal{C}$ must be to correct for --- or be impervious to --- such erasures.  If $k$ qubits are required to encode a message without error correction, and $l$ qubits are erased, $\mathcal{C}$ requires (\cite{almheiri:15} Eq. 3.22)
\begin{equation}
n \geq 2 l + k
\end{equation}
\noindent
qubits for error correction.  As remarked in \cite{almheiri:15}, this is an intuitive result: for each qubit that is erased, at least two others are required to reconstruct its state.}

{\rm More important for our purposes than the size of a codespace capable of error correction is its entanglement structure.   As shown in \cite{almheiri:15}, the codespace $\mathcal{C}$ can be protected against the erasure of a set $e$ of qubits only if (\cite{almheiri:15} Eq. 3.14)
\begin{equation}
\rho_{er}[\phi] = \rho_e[\phi] \rho_r[\phi]
\end{equation}
\noindent
where $r$ is the set of $k$ qubits added to the codespace to enable protection and all partial traces are over the entire $n$-qubit state $\phi$.  Separability between qubits in $e$ and those in $r$ clearly depends on the choice of basis for $\phi$ \cite{zanardi:00}.  Here we again see the dependence of the QECC on the availability of a classical channel that allows users of the code to coordinate choice of the same basis for their local $E$ and $D$ operators; i.e. the dependence of the QECC on LOCC at the level of the observers/users.  Quantum redundancy within the QECC thus depends on classical redundancy of basis choice, i.e. on gauge invariance \cite{addazi:21}.}

{\rm The final result of \cite{almheiri:15} that is of interest here is that the extent to which a qubit $q$ in $\mathcal{C}$ is protected from erasure functions, in an AdS/CFT context, as an effective radial coordinate for the AdS bulk -- qubits in the ``center'' of the bulk are easily protected, while qubits near the boundary are more vulnerable.  To see this in the generic representation of Diagram \eqref{locc-qecc}, we note that the entire AdS bulk is treated as the codespace in \cite{almheiri:15}, and that both the time evolution $U_{\mathcal{C}}$ and the encoding and decoding operators $E$ and $D$ are left implicit.  Let us consider the boundary CFT to be defined on $\mathscr{B}$ and consider each neighborhood of $\mathscr{B}$ to be a sector $S_i$ characterized by a local CFT observable $x_i$.  Decoding and encoding the state $|x_i \rangle$ on $S_i$ correspond, respectively, to obtaining a classical outcome value by measuring $|x_i \rangle$ using a local QRF $Q_i$ and to preparing $|x_i \rangle$, using $Q_i$, given such a classical value; measuring and preparing the state $|\uparrow \rangle$ of a single qubit provides an example.  The bulk neighborhood $\mathcal{C}_i$ of $S_i$ on which $|x_i \rangle$ is encoded is, as pointed out in  \cite{almheiri:15}, just the causal wedge $\mathscr{W}_{\mathcal{C}} [S_i]$.  Encoding $|x_i \rangle = |0 \rangle$ on $S_i$ erases the qubits in $\mathscr{W}_{\mathcal{C}} [S_i]$, but has negligible effect in the rest of $\mathcal{C}$, which encodes the combined values of $|x_j \rangle$ on $S_j$ for $j \neq i$.  We can, indeed, think of the qubits in $\mathscr{W}_{\mathcal{C}} [S_i]$ as implementing local operators $E_i$ and $D_i$ that act on the rest of $\mathcal{C}$, i.e. on $\mathcal{C} \setminus \mathscr{W}_{\mathcal{C}} [S_i]$.  In the language of quantum Darwinism, the $\mathscr{W}_{\mathcal{C}} [S_i]$ are the environmental fragments with which the local (to $S_i$) observers interact.}

Harlow \cite{harlow:18} has further explored this connection between proximity to the boundary and vulnerability to erasure by establishing an equivalence relation between erasure protection and the Ryu-Takayanagi generalization \cite{ryu:06} of the Hawking-Bekenstein black hole entropy.   If $X$ is a $d$-dimensional system with a $(d - 1)$-dimensional boundary $\partial X$ in a $(d+1)$-dimensional CFT, and $\gamma_X$ is a $d$-dimensional surface in AdS$_{(d+2)}$ with boundary $\partial X$, then \cite{ryu:06}, Eq. 1.5 defines an entanglement entropy:
\begin{equation} \label{ryu}
\mathcal{S}_X = \dfrac{\mathrm{Area ~of} ~\gamma_X}{4 G^{(d+2)}_N}\,,
\end{equation}
where $G^{(d+2)}$ is the gravitational constant in the bulk AdS$_{(d+2)}$. If $\Xi_X$ is the bulk region bounded by $\gamma_X \sqcup X$, the entanglement wedge is recovered:
\begin{equation} \label{wedge}
W[X] = D_{\rm{bulk}} [\Xi_X]\,,
\end{equation}
where $D_{\rm{bulk}}$ is the bulk domain of dependence.  We can, therefore, interpret Eq. \eqref{ryu} as relating the entanglement entropy of a system $X$ in the boundary CFT to the entanglement entropy of its corresponding wedge $W[X]$ in the bulk \cite{bain:20}. Erasing quantum information in one system, therefore, erases it in the other. Related is the approach adopted in \cite{lee:11}, in which redundancy and hence error-correction capacity on the boundary is again provided by entanglement within the bulk. Scrambling processes effectively remove redundancy from the bulk, and hence from any QECC implemented by the bulk \cite{SHM1, SHM2}. Thus scrambling in the bulk generates noise on the boundary.

\end{example}

\subsection{Spin networks and TQNNs} \label{STQECC}

We are now in a position to construct a generic model of spacetime as a representation of the classical redundancy enabled by a QECC.  Intuitively, what we will show is that the sectors $S_i$ of the manifold $\mathscr{B}$ can be considered neighborhoods with spatial coordinates that are, in a low-energy/low-resolution limit, related by a spatial metric.  We do this by showing that, given appropriate symmetries on $\mathscr{B}$, a TQFT on $\mathscr{B}$ can be represented by TQNNs (i.e., states of the spin-network basis on a Hilbert space) on $\mathscr{B}$ that, effectively, encodes topological connectivity as metric connectivity. This can be easily understood in the sense of dual Regge geometries. Since they are spin-network states, TQNNs represent simplicial Euclidean three-dimensional geometries on the boundaries.

We begin by adopting a spin-network approach.  Let us consider a graph $\Gamma$, embedded on some manifold $\mathcal{M}$. Let us colour its links $\gamma_{ij}$ with irreducible representations of a Lie group, and its nodes $n_i$ with intertwiner numbers $\iota$, labelling invariant tensorial representations of the same Lie group. The resulting states, supported on a generic graph $\Gamma$, are assumed to retain gauge-invariance at each node. These are the so called spin-network states, denoted as $|\Gamma, j, \iota \rangle$. In the holonomy representations, to each link $\gamma_{ij}\in \Gamma$ is associated a holonomy group element $h_{\gamma_{ij}}\equiv h_{ij}$. Spin-network states can be then represented as cylindric functionals $\Psi_{j,\iota}(h_{ij})$. We pick the Lie group to be SU$(2)$. At each one of the $N$ nodes $n$, one can apply a SU$(2)$ gauge transformation for each one of the $L$ links $\gamma$. Hence, the corresponding Hilbert space $\mathcal{H}_\Gamma$ of $|\Gamma, j, \iota \rangle$ can be represented as $\mathcal{H}_\Gamma=L^2(SU(2)^L/SU(2)^N)$, the space of square summable cylindrical functionals $\Psi(h_{ij})$ with SU$(2)$ gauge-invariance at each node $\Psi(h_{ij})=\Psi(g_{n_i}h_{ij} g^{-1}_{n_j})$.

The states of TQFT can be either represented as cylindrical functionals or spin-network states, and can be mapped into TQNNs \cite{fgm:22,mar:20,mar:22}. Representing such a TQFT as a map $\mathscr{T}: \mathscr{B}_{\rm in} \mapsto \mathscr{B}_{\rm out}$, any boundary $\mathscr{B}_{i}$ intermediate between $\mathscr{B}_{\rm in}$ and $\mathscr{B}_{\rm out}$ can be considered a ``layer'', the structure of which is characterized by the spin-network states/TQNNs embedded on it. The spin-network states encoded by any such boundary $\mathscr{B}_{i}$ can thus be considered an intermediate representation produced by partial processing of a state encoded on $\mathscr{B}_{\rm in}$, while a state encoded on $\mathscr{B}_{\rm out}$ is the result of a computation. From now on, we denote with ``TQNN'' not only the spin-network states on the boundaries $\mathscr{B}$ and the infinite intermediate layers --- that intertwine the $\mathscr{B}$s --- but the very TQFT amplitudes that instantiate the evolution among boundary states.

Provided $\mathscr{B}_{\rm in}$ and $\mathscr{B}_{\rm out}$ can be represented as manifolds tessellated into mutually-separable tiles as discussed above, 
a TQNN operating on $\mathscr{B}_{\rm in}$ will encode patterns of topological connectivity on $\mathscr{B}_{\rm in}$ at multiple scales on intermediate (virtual) boundaries, and encode an abstraction of the multi-scale pattern on $\mathscr{B}_{\rm out}$.  This is demonstrated in the case of the MNIST hand-written character dataset in \cite{mar:20}.  This process effectively abstracts discrete metric information, in the form of path lengths and tile-edges crossed, from the input graph $\Gamma$.  For scales large with respect to the tessellation scale, this metric information approaches a well-defined metric on $\mathscr{B}_{\rm in}$.  It is this process that defines effective spatial coordinates on $\mathscr{B}_{\rm in}$ as discussed below.  Mutual separability between tiles -- hence mutual independence of their associated observers -- renders these spatial coordinates effectively classical.

\begin{example}{\bf The HaPPY code}

{\rm We can represent a TQNN as a tensor network provided the above condition that $\mathscr{B}_{\rm in}$ and $\mathscr{B}_{\rm out}$ can be represented as manifolds tessellated into mutually-separable tiles is met \cite{ieee-2}.  Pastawski et al. \cite{pastawski:15} discuss the special case of ``perfect'' tensors, and show how these represent erasure-protection QECCs that are, in a natural sense, optimally efficient.  A perfect tensor is a tensor $T$ with $m$ indices, such that for any bipartition of the indices into sets of $n_1$ indices and $n_2$ indices, $n_1 + n_2 = m$, $T$ is a proportional to an isometric tensor from $V_1$ to $V_2$, where $V_i$ is the vector space spanned by the $n_i$ indices of $T$.  As pointed out in \cite{pastawski:15}, $T$ is perfect if $T$ is unitary and $n_1 = n_2 = n$.  In this case,  the tensor describes a pure state of $2n$ spins such that any set of $n$ spins is maximally entangled with the complementary set of $n$ spins.  It is this maximal entanglement condition that renders the ``HaPPY'' code constructed with such a tensor optimally efficient for erasure protection.}

\end{example}

\subsection{Constructing spacetime with a TQNN}

We review in this section how space-times can be connected to QECCs when TQNNs are taken into account. The crucial property is that TQNNs are by definition supported on spin-network states. The latter ones are in turn connected to simplicial geometries, and to the Regge calculus, respectively through the identification of dual links and nodes with simplexes of appropriate co-dimensions \cite{Rovelli_book}, and the identification of parameters entering the group elements with characteristic features of the discretized Regge geometries \cite{Rovelli:2010km}. 

A correspondence between QECC and space-time can now be recovered along the following lines:
\begin{itemize}
\item
We consider $n$-qudits that are spin-network states $|\Gamma, j, \iota \rangle$, supported on a graph $\Gamma$ and belonging to some $n$-dimensional Hilbert space $\mathcal{H}_\Gamma$. These states capture topological and metric information of TQNNs \cite{fgm:22, mar:20, mar:22};
\item
We associate the spin-network states to dual simplices that are simplicial Regge geometries discretizing boundaries of space-time manifolds --- these are the ``three-geometries'' $g^{(3)}$ taken into account on the boundaries;
\item
We let boundary states evolve through a functor that extends to the bulk. This is equivalent to considering the formal expression for the evolution of boundary three-geometries, which is realized summing over four-geometries $g^{(4)}$ in the bulk:
\begin{equation} \label{grathe}
\mathcal{W}[g^{(3)}_{\rm in }, g^{(3)}_{\rm out }]= \int_{g^{(3)}_{\rm in }}^{g^{(3)}_{\rm out }} \mathcal{D} g^{(4)} e^{\imath \mathcal{S}[g^{(4)}]}\,,
\end{equation}
with $\mathcal{D} g^{(4)}$ and $\mathcal{S}[g^{(4)}]$ respectively the measure and the specific (either gravitational or topological) action governing the dynamics of the four-geometries in the bulk;
\item
In the bulk, individuated by the 2-complex $\mathcal{C}$, the evolution of the nodes $n$ ---  we label nodes with a subscript $i$, i.e. $n_i$ --- describe edges $e$, while the evolution of links $\gamma$ --- we will denote as $\gamma_{ij}$ the links among two nodes $n_i$ and $n_j$ --- describes faces $f$; edges intersect one another at vertices $v$ --- see e.g. \cite{Rovelli_book, Bianchi:2010mw}. The cobordism between the space hypersurfaces encodes the space-time structure through: a face amplitude, which can be represented as a delta function over a product of holonomies $h_{v \gamma_{ij}}$ along the internal edges bounding the internal faces $f\in\mathcal{C}$; a vertex amplitude $\mathcal{W}_v(h_{v \gamma_{ij}})$, which is local in space-time and still depends on the holonomies $h_{v \gamma_{ij}}$, with a vertex label $v$ and a link label $\gamma_{ij}$. Without specifying $\mathcal{W}_v(h_{v \gamma_{ij}})$, which selects a particular theory (either gravitational or topological), one may realize the cobordism in terms of an integration over the bulk variables $h^{\rm bulk}_{v \gamma_{ij}}$, namely:
\begin{equation} \label{2cfun}
\mathcal{W}(h_{\gamma_{ij}})=\sum_\mathcal{C} \int d h^{\rm bulk}_{v \gamma_{ij}} \prod_{v\in \mathcal{C}} \mathcal{W}_v(h_{v \gamma_{ij}}) \prod_{f \in \mathcal{C}} \delta \!\left( \prod_{v \in \partial f}h_{v \gamma_{ij}} \right)\,.
\end{equation}
\end{itemize}

\subsection{QECC encoding for a BF TQFT in three dimensions}

We start by considering the theory in \eqref{grathe} to be a topological quantum field theory on a SU$(2)$ principal bundle and over a three dimensional space-time $\mathcal{M}_3$ base manifold, specified by the $BF$ action:
\begin{equation}
\mathcal{S}_{\rm BF}=\int_{\mathcal{M}_3} {\rm Tr}[B\wedge F]\,,
\end{equation}
with $F$ the field strength of the SU$(2)$ connection $A$, and $B$ a one-form valued in the $\mathfrak{su}(2)$ algebra, and where the trace ``Tr'' is over the adjoint indices of SU$(2)$. The $A$ and $B$ fields are conjugated variables, which individuate a canonical symplectic structure for the theory. The equations of motion of the theory are the Gau\ss ~constraint $d_A B=0$, generating SU$(2)$ gauge-transformations, and curvature constraint $F=0$ --- see e.g. the discussion in \cite{fgm:22}. The vertex amplitude of the theory, in the holonomy representation \cite{Bianchi:2010mw}, is finally expressed by:
\begin{equation} \label{BFsf}
\mathcal{W}^{\rm BF}_v(h_{v \gamma_{ij}})= \int \prod_{n_i=1}^4 d g_{n_i} \prod_{\gamma=1}^6 \delta(g_{n_i} h_{v \gamma_{ij}} g_{n_j}^{-1})\,.
\end{equation}
The holonomy representation is related to the ``spin and intertwiner'' representation by means of the Peter-Weyl transform, which allows us to decompose any spin-network cylindric functional as:
\begin{equation}
\Psi_{j_{\gamma_{ij}},\iota_{n_i} }(h_{ \gamma_{ij}})=\left(\bigotimes_n \iota_n \right)\cdot \left(\bigotimes_{\gamma_{ij}} D^{(j_{\gamma_{ij}})} (h_{\gamma_{ij}})\right)
\,,
\end{equation}
where $D^{(j)}$ are SU$(2)$ representation matrices and $\iota_n$ here denote  intertwining tensors.\\

In the spin and intertwiner representation, the classifier is then provided by the expression:
\begin{equation}\label{mael}
\mathcal{W}^{\rm BF}_v(j_{v \gamma}, \iota_n)=\langle \mathcal{W}^{\rm BF}_v | \Psi_{j_{v \gamma},\iota_{n} }
\rangle\,.
\end{equation}
Specifically, introducing the Racah $6j$-symbols $\{6j\}$, the three-dimensional BF action vertex amplitude, in the spin and intertwiner representation, acquires the expression:
\begin{equation}
\mathcal{W}^{\rm BF}_v(j_{v \gamma}, \iota_n)=
\frac{1}{\prod_{v \gamma} (2 j_{v \gamma} +1)}\, \{ 6j_{v \gamma}\}\,.
\end{equation}

The three-dimensional BF theory on SU(2) provides an example to define of a theory of TQNNs: TQNN states correspond to spin-network states defined on boundaries, and their evolution from a boundary to another is instantiated by means of Eq.~\eqref{BFsf}.

For $\partial \mathcal{C}=\Gamma$, we may now consider $\Gamma\equiv\Gamma_C=\Gamma_R\cup \Gamma_{R^c}$, which corresponds to the decomposition of boundary states $\Psi_{\Gamma_C}= \Psi_{\Gamma_R} \otimes \Psi_{\Gamma_{R^c}}$, and correspondingly to the decomposition of Hilbert spaces $\mathcal{H}_{\Gamma_C}=\mathcal{H}_{\Gamma_R}  \otimes \mathcal{H}_{\Gamma_{R^c}}$ on disjoint/complementary boundaries, in which graphs are embedded. The two-complex evolution (cobordism) among spin-network states living on the complementary boundaries is achieved by means of the matrix elements in Eq.~\eqref{2cfun}. It defines a tensor network that automatically encodes a unitary transform $U_R$, and hence are ``perfect'' in the sense of \cite{pastawski:15}.

Following \cite{bain:20, pastawski:15}, perfect tensor networks can be introduced defining a $k$-indexed state $\vert \wti{j_1 \cdots j_k} \rangle$. This latter corresponds to a $k$-codespace\footnote{Considering the Hilbert space of the holographic screen (functionally, the MB) $\mathscr{B}$, and following \cite{bain:20, almheiri:15, pastawski:15}, we may decompose a $n$-qudit product space into the product $\H^{(n)} = \H^{(m)}_R \otimes \H^{(n-m)}_{R^c}$, which involves products of $m$-qudits and $(n-m)$-qudits, supported respectively on the sets $R$ and $R^c$, $R^c$ denoting the complement of $R$. Logical qudits form a subspace $\H_C \subset \H^{(n)}$ called the {\em codespace}. The conditions for erasure-protection encoding on $\mathscr{B}$ are then met if and only if $R=R_1 \sqcup R_2$, with $\sqcup$ denoting disjoint union and $R_1$ and $R_2$ comprising $k$ qudits and $(m-k)$ qudits, respectively, such that $\vert \ovl{i}\rangle = \rm{U}_R (\vert i\rangle_{R_1} \otimes \vert \chi \rangle_{R_2 R^c})$, where $\vert \ovl{i} \rangle$ is an $n$-qudit basis of $\H_C$, $\vert i \rangle_R$ is a $k$-qudit state on $R_1$, $\vert \chi \rangle_{R_2R^c}$ is an $(n-k)$-qudit state on $R_2 \cup R^c$, and $\rm{U}_R$ is a unitary transformation acting non-trivially on $R$. Within the codespace $\H_C$ any $(n-m)$-qudit operator $O_{R^c}$ with support on $R^c$ is a multiple of the identity, i.e. $\langle \ovl{i} \vert O_{R^c} \vert \ovl{j} \rangle = c \delta_{ij}$, with $c$ a constant. This was proved in \cite{almheiri:15} for $\mathscr{B}$ in terms of the AdS-Rindler representation of a bulk field for which there are two commuting irreducible representations of the operator in question; hence by Schur's lemma the result follows. Note that throughout this description only Hilbert space theory is applied, thus $\mathscr{B}$ could just as well be the Hilbert space of a CS boundary, besides that of a CFT.
%
%
%
} states $\mu_1 \otimes \cdots \otimes \mu_k \in \H_C^{\otimes^k}$. Taking its inner product with the $n$-qudit basis state $\vert i_1 \cdots i_k \rangle$ then gives rise to a $(n+k)$-tensor:
\begin{equation}\label{qecc-4}
\langle i_1 \cdots i_n \vert \wti{j_1 \cdots j_k} \rangle = T_{i_1 \cdots i_n j_1 \cdots j_k}
\end{equation}
that encodes a unitary transform $\rm{U}_R$. Within the TQFT language, Eq.~\eqref{qecc-4} corresponds to Eq.~\eqref{mael}. This instantiates the `the HaPPY code' \cite{pastawski:15}. It is reversible, and  realizes an erasure-protection QECC encoding on one logical qudit in five physical qudits such that the former is protected against erasure of any two of the latter, where the encompassing framework for this interpretation is that of a discrete system on a lattice of negative curvature.

\subsection{CS theories on the boundaries and quantum groups}

Another instantiation of the perfect tensor-networks in Eq.~\eqref{qecc-4} that realizes the `HaPPY code' is achieved by taking into account another formulation of lower dimensional theory of gravity. A notable example of a BF theory on a three-dimensional manifold $\mathcal{M}_3$ is indeed provided by the Einstein-Hilbert action with cosmological constant $\Lambda$ on Euclidean space.  Within the TQFT literature, it is well known that the Turaev-Viro (TV) partition function \cite{TV} can be recast as a difference of two CS actions \cite{WittenJonesPolinomials}. At the same time, it is also well known that the Einstein-Hilbert action with cosmological constant on a three-dimensional manifold can be cast as a difference of two CS actions \cite{WittenQuantumGravity2+1D}. The Einstein-Hilbert action with vanishing cosmological constant on a three-dimensional manifold is expressed by the relation:
\begin{equation}
\label{EH2D}
\mathcal{S}= 2 \int_{\mathcal{M}_3} \!\!\!\! e^a \wedge (d\omega_a + \frac{1}{2}\epsilon_{abc}\, \omega^b\wedge\omega^c),
\end{equation}
where the frame $B$ one-form is denoted as $e^a=e^a_\mu \, dx^\mu $, the SU$(2)$ connection as $\omega^a=\frac{1}{2} \epsilon^{abc}\, \omega_{\mu bc }\, dx^\mu $, and by convention we set up $16\pi G=1$. \\

The action \eqref{EH2D} is invariant under local SU$(2)$ transformations, i.e.
\begin{eqnarray*}
\label{Tgauge}
&& \delta e^a= \epsilon^{abc}\, e_b \, \alpha_c\,, \\
&& \delta\omega^a= d\alpha^a + \epsilon^{abc}\, \omega_b\, \alpha_c\,,
\end{eqnarray*}
with $\alpha^a$ infinitesimal parameters of the SU$(2)$ transformation.
By varying the action \eqref{EH2D} with respect to the triad and the spin connection, we find the equations of motion, respectively:
\begin{eqnarray*}
&&T^a[e,\omega]= de^a + \epsilon_{abc} \omega^b\wedge e^c =0\,,\\
&& R^a[\omega]= d\omega^a + \frac{1}{2} \epsilon_{abc}\omega^b\wedge
\omega^c =0\,.
\end{eqnarray*}
The equation $T^a[e,\omega]=0$ is the torsion-free condition that determines $\omega$ in terms of $e$, while $R^a[\omega]=0$ implies that the connection $\omega$ is flat.  It is possible to check that the CS action \cite{TV} recasts as:
\begin{equation}
\mathcal{S}_{\rm CS}= \frac{k}{4\pi} \int_{\mathcal{M}_3}\!\! {\rm Tr}( A\wedge dA +
\frac{2}{3}A\wedge A\wedge A)\,,
\end{equation}
which is simply the action \eqref{EH2D}, where $A$ denotes the spin connection.
The action \eqref{EH2D} can be expressed, once the cosmological constant $\Lambda$ is introduced, as:
\begin{equation}
\label{frrr2}
\mathcal{S}= 2 \int_{\mathcal{M}_3} e^a \wedge (d\omega_a +
\frac{1}{2}\epsilon_{abc}\omega^b\wedge\omega^c)+ \Lambda \epsilon_{abc} e^a \wedge e^b \wedge e^c\,.
\end{equation}
For $\Lambda>0$, it is possible to express $\Lambda= 1/l^2$, and then reshuffle the Einstein-Hilbert action with cosmological constant as:
\begin{equation}
\mathcal{S}= \mathcal{S}_{\rm CS}[A^{(+)}]-\mathcal{S}_{CS}[A^{(-)}]\,,
\end{equation}
in which $A^{(\pm)a} = \omega^a \pm \frac{1}{l}e^a \,.$ \\

Within the BF theory formalism, the TQFT action with cosmological constant term reads:
\begin{equation}
\label{bffL}
\mathcal{S^{\rm BF-\Lambda}}= \int_{\mathcal{M}_3} {\rm Tr}(B \wedge F + \frac{\Lambda}{12}B \wedge B\wedge B)\,,
\end{equation}
where $F$ is a two-form that denotes the curvature of the $SU(2)$ connection $A$, $B$ is the one-form valued in $\mathfrak{su}(2)$, the algebra of $SU(2)$, and $\mathcal{M}_3$ is a three-dimensional orientable manifold.
The Turaev-Viro model previously reviewed is the discretization of the path integral of the action \eqref{bffL}. A detailed analysis of the equivalence between the quantization of the action \eqref{bffL} and the TV model has been provided in Ref.~\cite{Gresnigt:2022lwq}. Corresponding topological invariants of the model have been computed in \cite{kauffman:91}.\\

The relevance of the TV model is due to the fact that it implements the background independent quantization of the Einstein-Hilbert action with cosmological constant in three dimensions. A triangulated manifold, i.e. a triangulation $\Delta$ of $\mathcal{M}$, with edges labeled by irreducible representations of the quantum group $SU_q(2)$ --- equivalently, we may think of assigning representations of $SU_q(2)$ to the faces of the dual $2$-complex $\Delta^*$ ---  can be picked up. The amplitudes of the vertices of $\Delta^*$ are the $q$-deformed versions of $6j$-symbols $(6j)$ of $SU(2)$, which we call $q$-$6j$-symbols and denote with $(6j)_q$. The deformation parameter $q$ is typically assumed to be a root of unity, and can be defined as:
\begin{equation}
q= e^{\frac{2\pi \imath}{r}}\, ,
\end{equation}
in which $r$ is integer and such that $r\geq 3$ --- for further details about the quantum group $SU_q(2)$ and its $6j$-symbols see \cite{TV}.
The vacuum-vacuum transition amplitude of the theory is given by:
\begin{equation}\label{TVD}
\mathcal{Z}_{\rm TV}(\Delta) =  \eta^{2V} \sum_{j_e} \prod_e {\rm dim}_q (j_e) \prod_t (6j)_q\,,
\end{equation}
in which $V$ is the number of vertices in $\Delta$, the product index $t$ denotes the tetrahedra, ${\rm dim}_q(j)=[2j+1]_q$ represent the quantum dimension, and:
\begin{equation}
\eta = \dfrac{(q^{\frac{1}{2}} - q^{-\frac{1}{2})}}{\imath \sqrt{2k}}
\end{equation}
where $k=2\pi/\sqrt{\Lambda}$.
Eq.~\eqref{TVD} is independent of the triangulation $\Delta$, hence realizing a topological invariant of $\mathcal{M}$, namely:
$$\mathcal{Z}_{\rm TV}(\Delta)= \mathcal{Z}_{\rm TV}(\mathcal{M})\,.$$
Eq.~\eqref{TVD} rephrases the Ponzano-Regge partition function \cite{ponzano}, with the standard recoupling theory of $SU(2)$ substituted by the representations of $SU_q(2)$. In particular, the limit $\Lambda\rightarrow 0 $ entails the undeformed recoupling quantities of SU$(2)$, so that the Ponzano-Regge model is recovered. On the other hand, the deformation parameter $q\rightarrow 1$ when $r \rightarrow \infty$, which is equivalent to the limit $\Lambda \rightarrow 0$, by means of:
\begin{equation}
q= e^{\imath\frac{ \sqrt{\Lambda}}{2l_p}}\, ,
\end{equation}
where $l_p$ is the Planck length.

The amplitudes introduced in Ref.~\cite{Gresnigt:2022lwq}, which encode the recoupling theory of $SU_q$(2), generalize the results obtained in \eqref{mael} that involved the recoupling theory of $SU(2)$. These amplitudes provide the matrix elements for Eq.~\eqref{qecc-4} that are derived from the theory in Eq.~\eqref{bffL}.

\subsection{Marzuoli-Rasetti coding and TQFT as quantum simulator}

The universal representation of quantum structures has been advocated to have a crucial role in quantum information. In particular, the recoupling theory of the angular momentum provides the universal language necessary to develop a quantum simulator connected to topological quantum field theory \cite{Fred:02a,Fred:02b,Marzuoli:2002xq}. As summarized by Feynman \cite{Feyn:82}, a quantum simulator must fulfil locality of interactions, its computing elements should increase proportionally to the space-time volume it individuates, and time is simulated in terms of discrete computational steps. These are all properties fulfilled by spin-network states belonging to the Hilbert space of TQFT and realizing a holonomic quantum computing system \cite{Zan:99,Pa:00}. \\

Focusing on a quantum Turing machine $\mathfrak{M}$, satisfying specific axioms \cite{Manin:1999ic} and resulting as an instantiation of a quantum simulator, it is possible to identify the computational space of $\mathfrak{M}$ with spin-network states, and hence code information in terms of irreducible representations of SU(2). Suppose for instance to recover as building blocks of $\mathfrak{M}$ an ordered collection of $n+1$ mutually commuting angular momentum operators, which we denote as $\{ {\bf J}_l \}$, with $l=1,\dots, n+1$. Simultaneous eigenvectors of the square of ${\bf J}_l$ and its projection $\{J_z^l\}$ can be recovered, with eigenvalues respectively the half-integers $j_l$ and $m_l$, the latter ones appearing in integer steps $-j\leq m \leq j$. The two set operators $\{ {\bf J}_l \}$ and $\{J_z^l\}$ sum respectively in a total angular momentum ${\bf J}$, with eigenvalue $j$, and its component $J_z$, with eigenvalue $m$, the simultaneous diagonalization still implying integer steps $-j\leq m \leq j$. Compositions of couples of angular momenta $\{j_{l}, j_{l'}\}$ into generic $k_{ll'}$, in order to sum up to the total angular momentum $j$, involve $n-1$ intermediate angular momenta $k_s$, with $s=1, \dots, n-1$. The alphabet to perform an encoding of quantum information, given any pair $(n,j)$, is represented by the set of all possible binary coupling of the $n+1$ angular momenta $j_l$ and the intermediate $n-1$ angular momenta $k_s$. To each pair $(n,j)$ is associated a $(2j+1)$-dimensional Hilbert space $\mathcal{H}^j_n(k_1,\dots, k_{n-1})$, which are spanned by bases of the form:
\begin{equation} \label{bstate}
\{  | j_1, \dots, j_{n+1}; k_1, \dots, k_{n-1}; j, m \rangle \equiv | \mathfrak{b} \rangle \}\,.
\end{equation}

Composition of pairs of angular momenta can be pictorially represented with rooted binary trees, which on the other end are nothing but trivalent intertwiners. States of the type $ | \mathfrak{b} \rangle $ hence correspond to composition in a specific order of the rooted binary trees in a final tree, expressing the total angular momentum. Therefore, the tree is said to have root $j$ (corresponding to the total angular momentum of the system), with internal nodes to which are associated intertwiners and in which flow the intermediate angular momenta $k_1, \dots, k_{n-1}$, and terminal representations $j_1, \dots j_{n+1}$. Binary bracketings provide a representation of the states $ | \mathfrak{b} \rangle $ in terns of their combinatorial structure. Each of the assignments for $ | \mathfrak{b} \rangle $ correspond to a unique non-associative structure over the tensor product:

\begin{equation}
\mathcal{H}^{j_1} \bigotimes \dots  \bigotimes  \mathcal{H}^{j_{n+1}} \equiv {\rm span} \{ |j_1 m_1\rangle \otimes \cdots \otimes |j_{n+1} m_{n+1}\rangle  \}\,.
\end{equation}

Coding amounts to recovering generalized G\"{o}del numbers in bases associated to fields of ordered labels of intermediate angular momenta and coupling brackets.  A further notable feature of these structures is that they are naturally and intrinsically entangled. Finally, they provide the base to represent a specific class of ``binary'' TQNNs.\\

Operations are unitary quantum transformations that connect pairs of binary coupled states through recoupling coefficients, the $3nj$ symbols, representing (non-dynamical) probability amplitudes to measure a system initially described by the state $ | j_1, \dots, j_{n+1}; k_1, \dots, k_{n-1}; j, m \rangle $ into the state $| j_1, \dots, j_{n+1}; k_1', \dots, k_{n-1}'; j, m  \rangle$, i.e.

\begin{equation} \label{tnd}
\mathcal{M}_{3nj} \left[ \begin{array}{c} k_1, \cdots, k_{n-1}\\ k_1', \cdots, k_{n-1}' \end{array}\right]\equiv \langle  j_1, \dots, j_{n+1}; k_1', \dots, k_{n-1}'; j, m  | j_1, \dots, j_{n+1}; k_1, \dots, k_{n-1}; j, m \rangle \,.
\end{equation}

These are reduced matrix elements, recovered neglecting the total magnetic quantum number $m$ in accordance to the Wigner-Eckart theorem. They provide, as elements of the transfer matrices connecting any pair of states, the analog of the transition function of the quantum Turing Machine \cite{Manin:1999ic}. They also represent perfect tensor-networks, in the sense of \cite{pastawski:15}.\\

The computational space of $\mathfrak{M}$ is the graph $\Gamma$ whose vertices are identified by the specific combinatorial structure in \eqref{bstate}. Denoting with $(jj')_{k}$ a generic pairing of irreducible representations $j$ and $j'$ into $k$, a Racah transform $\mathcal{R}$ and a phase transform $\Phi$ can be defined for the states in \eqref{bstate} as
\begin{eqnarray} \label{loquaga}
&&\mathcal{R}: | \dots ((j_{l-1} j_{l})_{k_{s}}  j_{l+1})_{k_{s+2}} \dots \rangle  \qquad \longrightarrow  \qquad | \dots  (j_{l-1} (j_{l}  j_{l+1})_{k_{s+1}} )_{k_{s+2}} \dots \rangle\,,  \\
&&\Phi: | \dots (j_l j_{l-1}  ) \dots \rangle \qquad  \longrightarrow \qquad  | \dots (j_{l-1} j_l )  \dots \rangle
\,,
\end{eqnarray}
which can be respectively interpreted as rotation and twist of irreducible representations. The Biedenharn-Louck theorem ensures that $\mathcal{R}$ and $\Phi$ provide all possible transformations between pairs of binary couplings, at any $n$. We notice that the implementation of the tensor-networks dynamics in Eq.~\eqref{tnd}, namely the implementation of Eq.~\eqref{2cfun}, encodes the operations in Eq.~\eqref{loquaga}. \\

Marzuoli and Rasetti \cite{Marzuoli:2002xq} proposed a coding based on the operations that can be implemented through $\mathcal{R}$ and $\Phi$. The graph $\Gamma$ is then a twist-rotation graph with nodes associated to the computational states of the Turing Machine. In particular, the combinatorial structure of $\Gamma$ is determined by mathematical identities involving the $6j$ symbols:
\begin{itemize}
\item
the Biedenharn-Elliot identity, which generate pentagon plaquettes in $\Gamma$, i.e.
\begin{eqnarray}
&& \sum_w (-1)^{R+w}\, (2w+1) \left\{ \begin{array}{ccc} a& b & w \\ c & d& p \end{array} \right\}  \left\{ \begin{array}{ccc} c& d & w \\ e & f& q \end{array} \right\}  \left\{ \begin{array}{ccc} e& f & w \\ b & a& r \end{array} \right\} \nonumber\\
&& \qquad =  \left\{ \begin{array}{ccc} p& q & r \\ e & a& d \end{array} \right\} \left\{ \begin{array}{ccc} p& q & r \\ f & b & c \end{array} \right\}\,,
\end{eqnarray}
with $R=a+b+c+d+e+f+p+q+r$;
\item

\begin{equation}
\sum_w (-1)^{p+q+w}\, (2w+1) \left \{ \begin{array}{ccc} a& b & w \\ c & d& p \end{array} \right \} \left \{ \begin{array}{ccc} a& b & w \\ d & c& q \end{array} \right \} = \left \{ \begin{array}{ccc} a& c & q \\ b & d& p \end{array} \right \}\,.
\end{equation}

\end{itemize}

The computational power of a quantum computer modeled within this framework increases with the volume of the Hilbert space of the theory. For the specific topology of graphs considered in Ref.~\cite{Marzuoli:2002xq}, characterized by compositions of binary trees, the order of $\Gamma$, namely the number of vertices as a function of the number of irreducible representations assigned to the links of the $\Gamma$, increases factorially as $(2n-1)!$, hence for large $n$ as $n^n$. At the same time, the diameter of $\Gamma$ grows as $n \log n$, thus like the logarithm of its order, providing an upper bound for the time-length, namely the number of steps, that $\mathfrak{M}$ can perform.\\

The universality is ensured by the fact that unitary transformation instantiated by an operation of $\mathfrak{M}$ can be cast in finite sequences of operations on $\Gamma$. On the other hand, being also locality of the interactions --- as it appears from the bracketing structures ---  and discreteness of the evolution in the (time-lapse) steps fulfilled, $\mathfrak{M}$ identifies as a quantum simulator, precisely in the sense specified by Feynman \cite{Feyn:82}.

In particular, the combinatorial structure of $\Gamma$ and the number of computational steps enable to simulate the dynamical evolution in time lapse from initial states $| {\rm in} \rangle$ to final states $| {\rm out} \rangle$. There exists therefore an inherent discreteness in the quantum simulator structure, which shares the very same topological origin of its entanglement structure. On the other hand, exactly as in a classical Turing Machine, in a quantum Turing machine computation can be regarded as a map from input data to output data, instantiated by the unitary transformations $\mathcal{R}$ and $\Phi$. These are, on the other hand, implemented through a cobordism that instantiates Eq.~\eqref{2cfun}. The structure of the computation in $\mathfrak{M}$ can be then seen as a generalisation of the Boolean scheme, with the coding alphabet consisting of all the irreps $j_l$ labelling the coupled momenta, the intermediate $k_s$ momenta and the bracketing structure. An algorithm can be then specified by an ordered sequence of gates that are local alterations induced by either $\mathcal{R}$ or $\Phi$ (of the labels characterising the states). The amplitude of the unitary operator $\mathcal{U}$ that can be associated to each local alteration singles out an Hamiltonian operator $H ( \mathfrak{b}_{\alpha+1},  \mathfrak{b}_{\alpha})$ in the $\alpha$-th step:
\begin{equation}
 \langle \mathfrak{b}_{\alpha+1} | \mathcal{U} | \mathfrak{b}_{\alpha}  \rangle = e^{\imath H ( \mathfrak{b}_{\alpha+1},  \mathfrak{b}_{\alpha}) \tau} \,,
\end{equation}
while the whole sequence of local alterations are considered to complete the program and find the overall evolution of $| {\rm in} \rangle$ states into $| {\rm out} \rangle$ states, namely:
\begin{equation}
\langle {\rm out}| \mathcal{U} | {\rm in}  \rangle \equiv \prod_{\alpha=0}^{N-1} \langle \mathfrak{b}_{\alpha+1} | \mathcal{U} | \mathfrak{b}_{\alpha}  \rangle\,,
\end{equation}
where $\{| \mathfrak{b}_{\alpha}  \rangle\}$, with ordering $\alpha=0,\dots, N$, denotes a sequence of states such that $| \mathfrak{b}_{0}  \rangle = |{\rm in}\rangle$ and $| \mathfrak{b}_{N}  \rangle = |{\rm out}\rangle$.

Entering a longstanding and still unsolved dispute \cite{Pal:00}, in \cite{Marzuoli:2002xq} it was argued that spin-networks could be used as a quantum simulator supporting quantum algorithms based on TQFTs, and that these algorithms could enable recovering a path with minimized length between two given vertices of $\Gamma$, conjectured to be a {\bf NP}-c problem. This perspective was supported by the observation \cite{Sleat:98} that the maximum distance between any pair of binary trees, enclosing $N$ internal nodes, is at most linear in $N$.

\subsection{TQFT and the search for quantum algorithms}
\noindent
In \cite{Garnerone:2006df}, the authors provided quantum algorithms that implement efficiently approximations of colored Jones polynomials. Considering the SU(2)-symmetric CS action on $S^3$, namely:
\begin{equation}
\mathcal{S}_{\rm CS}(A) = \frac{k}{4 \pi} \int_{S^3} {\rm Tr}[A\wedge dA +\frac{2}{3} A\wedge A\wedge A]\,,
\end{equation}
with $A$ connection one-form valued in the $\mathfrak{su}(2)$ algebra, and the coupling constant $k$ defining the level of the CS theory.\\

The partition function of the theory $\mathcal{Z}$ is expressed by:
\begin{equation}
\mathcal{Z}_{\rm CS}[S^3,k] = \int [\mathcal{D}A] e^{\frac{\imath k}{4 \pi} \mathcal{S}_{\rm CS}(A)  } \,,
\end{equation}
which constrains $k$ to be a positive integer for gauge-invariant quantization. For any manifold $\mathcal{M}$ such that $\partial \mathcal{M} =0$, $\mathcal{Z}_{\rm CS}[\mathcal{M},k]$ defines a topological global invariant, due to the finiteness of the space of solutions of the CS theory.\\

Gauge-invariant observables of the theory cast as expectation values of Wilson-line operators supported on oriented knots and links embedded on three-manifold $\mathcal{M}_{\rm 3D}$. Wilson loops of the theory are group elements of SU(2) whose irreducible representations are labelled semi-integer spin numbers. The irreducible representations of $U_q(\mathfrak{su}(2))$, the $q$-deformation of the enveloping algebra of $\mathfrak{su}(2)$, often denoted as $SU_q(2)$, encode a relation between the parameter $q$ and the level $k$ of the CS action, namely $q=\exp(\frac{-2\imath \pi}{k+2})$. Exploiting quantization techniques for TQFT that encode the imposition at the quantum level of the curvature constraint on the quantized kinematical Hilbert space of the theory, and addressing specifically Euclidean SU(2)-symmetric gravity in 2+1D with positive cosmological constant, corresponding to the difference of two SU(2)-symmetric CS actions, in \cite{Gresnigt:2022lwq} it was shown the $q$-deformation of the enveloping algebra of $\mathfrak{su}(2)$, which leads to the emergence of the quantum group $SU_q(2)$, arises from the imposition of the curvature constraint with cosmological constant on Wilson loops.\\

A Wilson loop operator can be associated to a knot $\mathcal{K}\subset S^3$ that can be considered, for a fixed root of unity $q$, as the trace of the irreducible representation of spin $j$ of the holonomy of the connection one-form $A$, i.e.
\begin{equation}
W_j[\mathcal{K};q]= {\rm Tr}_j [ P e^{\oint_\mathcal{K} A} ]\,,
\end{equation}
where $P$ denotes path ordering. \\

Considering a link $\gamma$, decomposed in a collection of knots $\mathcal{K}_l$ with $l=1,\dots s$, composite Wilson loop operators can be expressed as:
\begin{equation}
W_{j_1, j_2 \dots j_s}[\gamma;q]= \prod_{l=1}^{s} W_{j_l}[\mathcal{K}_l;q]\,,
\end{equation}
and their expectation value calculated as:
\begin{equation}
\mathcal{W}_{j_1, j_2 \dots j_s}[\gamma;q]=
\frac{ \int [\mathcal{D}A]  \, {W}_{j_1, j_2 \dots j_s}[\gamma;q]
e^{\frac{\imath k}{4 \pi} \mathcal{S}_{\rm CS}}
}{\int [\mathcal{D}A]  e^{\frac{\imath k}{4 \pi} \mathcal{S}_{\rm CS}} }
\,.
\end{equation}
This  expression has been shown in \cite{Re:91,Kir:91,Li:97} to correspond to colored Jones polynomials, which for the specific choice of a fundamental representation reduces to the Jones polynomial\footnote{Colored link invariants are more efficient than standard Jones polynomials in detecting knots \cite{Ramadevi:1993hu}. On the other hand, introducing the $SU(2)_k$ boundary Wess--Zumino--Witten conformal field theory into the Chern–Simons action, enables developing a theory of colored oriented braids, thanks to the duality properties of the conformal blocks for the correlators --- see \cite{Kaul:1993hb}.} introduced in \cite{Jo:85}. Specifically, for a colored Jones polynomial $J_{j_1, j_2 \dots j_s}(\gamma;q)$, it holds that:
\begin{equation}
\mathcal{W}_{j_1, j_2 \dots j_s}[\gamma;q] =\left( \frac{q^{-3 w(\gamma)}}{q^{\frac{1}{2}} - q^{- \frac{1}{2}} } \right) J_{j_1, j_2 \dots j_s}(\gamma;q)\,,
\end{equation}
where $w(\gamma)$ is associated to the planar diagram $D(\gamma)$ of the oriented link $\gamma$, defined by the sum over all the crossing points $w(\gamma)=\sum_p \varepsilon(p)$ with $\varepsilon(p)=\pm1$, evaluated from the link diagram in virtue of counting arguments. \\

Given these definitions, it has been shown in \cite{Garnerone:2006df} that is possible to develop quantum algorithms, along the directions of \cite{Marzuoli:2002xq}, that are able to approximate the colored Jones polynomials. More precisely, it has been shown that, given a colored braid $b\in B_{2m}$ ($B_n$ denoting the braid group acting on $n$ fundamental representations) with length $\ell$, coloring ${\bf c}$, positive integer level $k$ and real $\delta>0$, it is possible to sample from a random variable $Z$, which is an additive approximation of the absolute value of the colored Jones polynomial of the plat closure of $b$, evaluated at $q=\exp(\frac{-2\imath \pi}{k+2})$.

This result confirms, in a different language, the representation of arbitrary sequences of measurements by TQFTs demonstrated in \cite{fgm:22}.  An algorithm is a sequence of steps, that is, a sequence of executions of operators that are not entangled in time.  This is precisely what a sequence of measurements is.  Such sequences approximate TQFTs which, as functors, are arbitrarily associative in time.\\

Finally, new quantum algorithms from the perspective of TQNN have been delved in \cite{LZM} and are being applied in \cite{LFFZM}. The specific tessellation that has been chosen to develop in \cite{LZM} the new quantum algorithms correspond to that one of topological bi-dimensional materials. Graphs are then supported on honeycomb hexagonal lattices, which have been proven to be useful in describing changes of topological phase in matter. This point will be clarified in the next subsection.

\subsection{Fault-tolerant storage of quantum information and topological phases}\label{fault-tolerant}

\noindent
The connection between topological gauge theories in three-dimensions and two-dimensional integrable lattice models has been widely investigated in the literature, starting from a seminal analysis by Witten \cite{Wit_GTILM:89}. This scenario then flourished again over the last decade, due to recent experimental and theoretical attempts \cite{DasSar:07,Kit:03} to develop fault-tolerant quantum computers, functioning on non-abelian quantum Hall states. Several models hitherto developed at this purpose hinge on the Chern--Simons--Witten theory. \\

Focusing on the Turaev-Viro model, which we review in details in the next section, Sec.~\eqref{m-theory}, its extension to a pair $(\mathcal{M}, \partial\mathcal{M})$, in order to account for three-manifolds with two-dimensional boundaries and deal with observables expressed in terms of embedded links and (ribbon) graphs, was achieved in two different ways in \cite{Kar:92,Kar:93} and \cite{Car:00}. Observables expressed in terms of colored graphs and satisfying braiding relations, have been studied in \cite{Kar:92,Kar:93,Be:95}, for oriented triangulated compact manifold $(\mathcal{M}, \partial\mathcal{M})$.\\

Hinging on this theoretical description, in \cite{Kadar:2009fs} it was emphasized that doubled topological phases, introduced by Kitaev \cite{Kit:03} and by Levin and Wen \cite{Wen:95,Levin:2004mi}, being supported on two-dimensional lattices, correspond to the Hamiltonian versions of three-dimensional TQFTs described by the Turaev-Viro state sum model. This observation enables to derive string-nets models \cite{Levin:2004mi} from TQFTs in the continuum. Specifically, the equivalence among the Kitaev and the Levin-Wen model, from one side, and the Turaev-Viro model has been proven for a honeycomb lattice and a finite group, resorting to duality transformation that connect the group algebra, and to the spin-network basis of lattice gauge theory. In \cite{Kadar:2009fs}  the authors further analyzed ribbon operators, with the aim of describing excitations in this class of models and providing a geometrical interpretation.

Lattice models developed in \cite{Kit:03,Levin:2004mi} implemented the description of microscopic degrees of freedom in terms of emerging topological phases, describing ground-states and quasi-particle excitations that are insensitive to local disturbances. Implementations of the string-nets framework in quantum computation were addressed in \cite{DasSar:07}, while their continuum limit has been observed to be related to the spin-network simulator \cite{Marzuoli:2002xq,Garnerone:2006df}. At the same time, entanglement entropy renormalization schemes have been applied to to the study of the ground states \cite{Agu:07,Koe:08}, and a tensor-network representation has been constructed in Ref.~\cite{Bue:08}. The entanglement properties were then studied in \cite{Ham:05,Levi:05}.

At the base of the definition of string-nets are the very same structures serving as algebraic data in TQFT. In \cite{Levin:2004mi} Levin and Wen considered a two-dimensional honeycomb lattice model over a surface $\Sigma$. Denoting the honeycomb lattice with $\Gamma$, which is a graph embedded on $\Sigma$, the possible states of the theory are those ones with any set of compatible representations that span the Hilbert space $\mathcal{H}_{\Gamma}$ at fixed graph $\Gamma$. To keep the notation close to the Turaev-Viro model, we may color $\Gamma$ with irreducible representations of $SU_q(2)$.
Within the lattice gauge theory framework \cite{Ban:77,Ko:75,Wen:03}, all de-confined theories admit an expression in terms of string-net condensates, where the strings represent the electric flux lines. The Hamiltonian acquires the form
\begin{equation}
H=-\sum_v Q_v - \sum_p B_p\,,
\end{equation}
where $Q_v$ and $B_p$ are two mutually commuting constraints operators. The ``electric-charge'' constraint operator $Q_v$ selects the configurations with vanishing charge, at each vertex $v\in\Gamma$, in terms of the fusion coefficients $N_{ijk}$ between irreducible representations. For $i$, $j$ and $k$ irreducible representations decorating the edges adjacent to a vertex $v$, the electric-charge constraint can be expressed by
\begin{equation}
Q_v=N_{ijk}\,,
\end{equation}
with $N_{ijk}$ such that $i\otimes j= \sum_k N_{ijk} \,k$. This can be thought as the discretized version of the gauge constraint, imposing the compatibility at the vertices $v$ among the irreducible representations. The ``magnetic flux'' constraint operator $B_p$ is defined on each hexagonal plaquette $p\in\Gamma$ and favors the states with no flux, providing the dynamics of the string-net configurations. It can be thought as the smeared version of the projection onto $\Sigma$ of the curvature constraint, and can be recast as a sum over irreducible representations $s$
\begin{equation}
B_p= \sum_s a_s B^s_p\,,
\end{equation}
where $a_s$ are real coefficients depending on $s$. The generic plaquette operator $B_p^s$ acts on each plaquette-component of the quantum states supported on the generic plaquette $p$, namely on
\begin{equation}
|p\rangle=|p[a,b,c,d,e,f;g_{ab}, h_{bc},i_{cd},j_{de},k_{ef},l_{fa}] \rangle\,,
\end{equation}
with irreps $a,b,c,d,e,f$ outgoing the hexagonal ring, and $g_{ab}, h_{bc},i_{cd},j_{de},k_{ef},l_{fa}$ irreps on the hexagonal ring --- $g_{ab}$ stands for the irrep on the plaquette $p$ between the two intertwiners in which the irreps $a$ and $b$ flow, and so on ---  according to
\begin{eqnarray}
&&B_p^s |p[a,b,c,d,e,f;g_{ab}, h_{bc},i_{cd},j_{de},k_{ef},l_{fa}] \rangle= \\
&&\!\!\!\!\!\!\!\!\!\!\sum\limits_{g',h',i',j',k',l'}  F^{bg*h}_{s*h'g*'} F^{ch*i}_{s*i'h*'} F^{di*j}_{s*j'i*'} F^{ej*k}_{s*k'j*'}  F^{fk*l}_{s*l'k*'} F^{al*g}_{s*g'l*'} \, \, 
|p[a,b,c,d,e,f;g_{ab}', h_{bc}',i_{cd}',j_{de}',k_{ef}',l_{fa}']  \rangle \nonumber\,,
\end{eqnarray}
where $j*$ represents the irrep dual to $j$, the numbers $F^{ijk}_{lmn}$ are 6-j symbols renormalized with respect to the Wigner 6-j symbols by
\begin{equation}
F^{ijk}_{lmn}=(2n+1) \left\{ \begin{array}{ccc} i & j & m\\ k & l & m \end{array} \right\}\,,
\end{equation}
which, accordingly, satisfy the Biedernharn-Elliot identity
\begin{equation}
\sum_{n=0}^{N} F^{mlq}_{kp*n}  F^{jip}_{mns*} F^{js*n}_{lkr*} = F^{jip}_{q*kr*} F^{r i q*}_{mls*}\,.
\end{equation}
The equivalence between the string-net model introduced by Levin and Wen and the Turaev-Viro TQFT, first notably shown in \cite{Kadar:2009fs,Koe:10}, and then delved in \cite{Kir:11}, can be exposed inspecting the string-net states on $\Sigma$ that correspond to boundary states for the TQFT. It has been then proven in \cite{Kadar:2008hq} that the transition amplitude generated by the magnetic flux constraint between two string-nets $\Gamma_{j}$ and $\Gamma_{j'}$ is connected to the Turaev-Viro amplitude by the relation
\begin{equation}
\langle \Gamma_{j}| \prod_p B_p |\Gamma_{j'}\rangle=
\mathcal{Z}_{\rm TV}[\Sigma\times [0,1], \tilde{\Gamma}_{j} , \tilde{\Gamma}_{j'} ]\,,
\end{equation}
where the Turaev-Viro invariant is calculated on the three-dimensional manifold  $\mathcal{M}_3=\Sigma\times[0,1]$, on which fixed triangulations on the two boundaries have been considered that are graphs $\tilde{\Gamma}_{j}$ and $\tilde{\Gamma}_{j'}$ dual to the graphs ${\Gamma}_{j}$ and ${\Gamma}_{j'}$, and labels are inherited from these latter ones.\\

Therefore string-nets, the transition amplitudes among which are connected to topological invariants in TQFT, and which in general can be represented as boundary spin-network states, encode along the lines of \S \ref{STQECC} the same information stored in $n$-qudits, which can be used to reconstruct boundary-geometries. In this sense, \emph{the Levin-Wen string-net model is a specific example of QECC--space-time}.

From the broader perspective that we have emphasized here, it becomes clear why such models should produce QECCs interpretable as spacetimes.  As emphasized in \cite{addazi:21}, spacetimes enforce separability.  From an error-correction perspective, separability is a resource: it is the resource that allows one to keep encoded values distinct from each other.  Hence any QECC must, effectively, encode a structure with the properties of a spacetime. \footnote{We expect that the discussion of \S\ref{fault-tolerant} bears relevance to recent advances such as in \cite{bravyi:23} where it is shown that certain QECC protocols induce fault-tolerant quantum memory based upon low-density parity-check coding.}

\subsection{The holomorphic representations in the semiclassical limit: From QECC back to LOCC}

To inspect the large spin (semi-classical) limit of TQNNs, we can resort to spin-network states whose nodes are labeled by coherent intertwiners \cite{liv:07,con:09,frei:10}. Coherent intertwiners that belong to the invariant tensor space Inv$(\otimes^L_{l=1} \mathcal{H}^{(j_l)})$ are labeled by $L$  unit-vectors $\vec{n}_l$ that fulfil the closure condition $\sum_l j_l \vec{n}_l$. Coherent intertwiners are expressed in terms of the Bloch SU$(2)$ coherent states $|j_l, \vec{n}_l \rangle$ as:
\begin{equation}
\Phi(\vec{n}_l)=\int_{\rm SU(2)} dg \bigotimes_{l=1}^L D^{(j_l)}(g) |j_l, \vec{n}_l \rangle\,,
\end{equation}
hence introducing the coherent spin-network states:
\begin{equation}
\Psi_{j_{\gamma_{ij}}, \vec{n}_i , \vec{n}_j' }(h_{\gamma_{ij}})=\left(\bigotimes_n \Phi(\vec{n}_i) \right)\cdot \left(\bigotimes_{\gamma_{ij}} D^{(j_{\gamma_{ij}})} (h_{\gamma_{ij}})\right)
\,,
\end{equation}
which form an over-complete basis of the Hilbert space $\mathcal{H}_\Gamma$ and enable a ``spin and normal representation'' of the vertex amplitude of the classifier:
\begin{equation}
\mathcal{W}_v({j_{\gamma_{ij}}, \vec{n}_i , \vec{n}_j' })=
\langle \mathcal{W}_v | \Psi_{j_{\gamma_{ij}}, \vec{n}_i , \vec{n}_j' } \rangle
\,.
\end{equation}
Leveraging coherent spin-networks to find a novel representation of the functorial evolution $\mathcal{W}$ of TQNNs, one can resort to the expression in terms of $H_{\gamma_{ij}}\in$ SL$(2, \mathbb{C})$, which exploits the analytic continuation to SL$(2, \mathbb{C})$ of the heat-kernel on SU$(2)$ $\mathcal{K}_{t}=\sum_j (2j+1) e^{-j(j+1)(t/2)} {\rm Tr}D^{(j)}(h)$. This is parametrized by $t_{\gamma_{ij}}$, namely \cite{bian:10}:
\begin{equation}
\Psi_{H_{\gamma_{ij}}}(h_{\gamma_{ij}})= \int \prod_n dg_n
\prod_{\gamma_{ij}} \mathcal{K}_{t_{\gamma_{ij}}}(g_{n_i} h_{\gamma_{ij}} g_{n_j'}^{-1} H^{-1}_{\gamma_{ij}})\,.
\end{equation}
Consequently, the coherent vertex amplitude can be expressed as:
\begin{equation}
\mathcal{W}_v(H_{\gamma_{ij} })= \langle \mathcal{W}_v | \Psi_{H_{\gamma_{ij} }} \rangle\,,
\end{equation}
which for the BF topological TQFT that corresponds to the Einstein-Hilbert action in three dimensions, reads:
\begin{equation}
\mathcal{W}^{\rm BF}_v(H_{\gamma_{ij} })=
 \int \prod_n dg_n
\prod_{\gamma_{ij}} \mathcal{K}_{t_{\gamma_{ij}}}((g_{n_i} H_{\gamma_{ij}} g_{n_j'}^{-1} )^{-1})\,.
\end{equation}
From the decomposition of the SL$(2,\mathbb{C})$ group elements, it is possible to express $H_{\gamma_{ij}}$ as:
\begin{equation}
H_{\gamma_{ij}}= h_{\vec{n}_i} e^{-\imath z_{\gamma_{ij}} (\sigma_3/2)}  h^{-1}_{\vec{n}_j'}\,,
\end{equation}
with $\sigma_3$ generator of the $\mathfrak{su}(2)$ algebra and:
\begin{equation}
z_{\gamma_{ij}}=\xi_{\gamma_{ij}}+ \imath a_{\gamma_{ij}} t_{\gamma_{ij}}\,.
\end{equation}
The simplicial space-time interpretation of $\xi_{\gamma_{ij}}$ is that of an extrinsic angle, associated to each link $\gamma_{ij}$ of the graph $\Gamma$ and coding a rotation generated by the extrinsic curvature; while $a_{\gamma_{ij}}$ is interpreted as the area of the surface, dual to the link $\gamma_{ij}$, shared by the cells (simplicial or polyhedral) associated to the nodes $n_i$ and $n_j$. The cells are the building blocks of a (cellular) decomposition of the boundary $\Sigma$ of $\mathcal{M}$; they can be singularly equipped with a three-dimensional Euclidean geometry, and then characterized by fixed shapes, with associated area and normals. The SU$(2)$ group elements $h_{\vec{n}_i}$ and $h^{-1}_{\vec{n}_j'}$, are individuated by the two normals $\vec{n}_i$ and $\vec{n}_j'$, as seen from each node, respectively $n_i$ and $n_j$. Congruent faces give rise to simplicial geometries, associated to Regge geometries.\\

For macroscopic values of $a_{\gamma_{ij}}$, namely $a_{\gamma_{ij}}\!>\!\!>\!1$, the representation matrices provide gaussian weights to each link ${\gamma_{ij}}\in \Gamma$, namely:
\begin{equation}
e^{j_{\gamma_{ij}}(j_{\gamma_{ij}}+1)(t/2)} D^{j_{\gamma_{ij}}}(H_{\gamma_{ij}}) \simeq e^{-\imath \xi_{\gamma_{ij}} j_{\gamma_{ij}} } \, e^{-(j_{\gamma_{ij}} - a_{\gamma_{ij}})^2 \frac{t}{2} + a_{\gamma_{ij}}^2 \frac{t}{2} } |j_{\gamma_{ij}}, \vec{n}_{i} \rangle \langle j_{\gamma_{ij}}, - {\vec{n}_j'} |\,,
\end{equation}
expressed in terms of the Bloch coherent states. This automatically selects large irreducible representations, the ones not suppressed by the gaussian weights, and consequently induces a specific topology, the one associated to the non-vanishing parameters $a_{\gamma_{ij}}$ that are assigned to each link $\gamma_{ij}\in \Gamma$. The process of topology selection generated by the limit $a_{\gamma_{ij}}\!>\!\!>\!1$ is reminiscent of the formal semi-classical $\hbar \rightarrow 0$ limit. This enables classical path selection by minimizing the classical action at complex exponential of the partition function. Similarly, since the inverse of $a_{\gamma_{ij}}$ plays the role of $\hbar$, the limit $a_{\gamma_{ij}}\!>\!\!>\!1$ implements quantum annealing on TQNNs:
\begin{equation}
\mathcal{W}^{\rm BF}\simeq e^{\imath \mathcal{S}_{\rm Regge}(j_{\gamma_{ij}}, \vec{n}_i, \vec{n}_j' ) } + e^{-\imath \mathcal{S}_{\rm Regge}(j_{\gamma_{ij}}, \vec{n}_i, \vec{n}_j' ) }
\end{equation}
for large $j_{\gamma_{ij}}$ and with $\mathcal{S}_{\rm Regge}$ denoting the Regge action. This process then singles out the boundary of the classifier architecture, selecting a fixed topology \cite{mar:22}.

As discussed in Part I \S 3, what is essential to the success of LOCC -- and hence for the utility of any QECC -- is the ability of some set of observers to consider one of their communication channels to be classical.  It is this classical channel that enables agreement about QRFs and hence measurement-basis choices as discussed in Part I and above.  In practice, classical communication is achieved through the use of macroscopic media to encode the information to be communicated.  All such media can be regarded as encoding information in stable geometric structures (i.e. shapes), with sound waves in air or characters printed on paper as canonical examples.  The primary requirement for stability of such structures is decoherence, i.e. separability from the surrounding environment, including both the physical implementation of the channel and the communicating agents.  A fixed, effectively-classical embedding geometry that imposes a spatial separation between the observers, and between the observers and their communication media, provides a mechanism to enforce separability \cite{addazi:21}.  Any such geometry requires a fixed topology, on which the notion of a classical path is meaningful.  Hence the classical limit described here enables the decoherence condition required for the classical-communication component of LOCC.

\section{Topological M-theory as a QECC} \label{m-theory}

In this section we investigate at the theoretical level how QECC can be realized in higher (space-time) dimensions than three, and consider the relation of the TQFTs (and their extensions) involved in the construction to theories of gravity in four space-time dimensions, and to M-theory, in higher than four dimensions.\\

Topological M-theory, provides a theoretically rich extension to extra dimensions of a relevant class of TQFT models. TQFT models, indeed, can be shown to correspond to QECCs. In this section, addressing this topic along the lines drawn in \S \ref{STQECC}, we extend the correspondence of TQFTs to QECCs to a correspondence to relevant models in Topological M-theory.  This provides us with an alternative ``picture'' of spacetimes as resources for separability, and hence as platforms for discrete, sequential processes. At the same time, this extends the analogue gravity perspective we have been driven by to the inclusion of a notable class of models connected to String Theory.

\subsection{Topological M-theory as a unifying model of gravity}

A notion of topological M-theory in $7$ dimensions was introduced in \cite{Dijkgraaf:2004te}, the classical solutions of which involve $G_2$ holonomy metrics. Remarkably, the theory provides a unification of the form-theories of gravity in various dimensions, being expressed in terms of a topological action for a Hitchin $3$-form gauge field. Their dimensional reductions encode $6$-dimensional topological A and B models, shedding light on their mutual S-duality, as well as self-dual Loop Quantum Gravity in $4$-dimensions and CS gravity in $3$ dimensions. \\

To unveil the relation between Superstring Theory and M-theory, one could consider that topological strings on Calabi-Yau $3$-folds and topological strings embedded into superstrings are connected, and then that there exist dualities for superstrings, geometrically explained in M-theory, arising from dualities in topological theories, with similar geometric explanation in topological M-theory. A natural definition of M-theory hence introduces an extra dimension, which is relative to the topological string, and enables one to state that a topological string theory on a Calabi-Yau manifold M is equivalent to an M-theory on M$\times S^1$, where one expects that the radius of $S^1$ can be mapped to the coupling constant of the topological string. If instead of assuming $S^1$ at constant size, one allows the radius to vary, supersymmetry-preserving manifolds in $7$ dimensions can be chosen that are geometric in the class of $G_2$ holonomy spaces. A M-theory on a $G_2$ holonomy manifold $X$ with a U$(1)$ action is then considered, so that the topological M-theory on X is defined as equivalent to A-model topological strings on X/U(1) that are provided with Lagrangian D-branes inserted at the points where the circle fibration degenerates. As the topological A-model can be understood as a classical theory of K\"{a}hler metrics with quantum corrections provided by strings wrapping holomorphic cycles, topological M-theory can be seen classically as a theory of $G_2$ holonomy metrics, with quantum corrections provided by membranes wrapping associative 3-cycles. The coupling between the membrane and the metric can be then straightforwardly inspected. It is natural to identify the $3$-form $\Phi$, which $G_2$ is equipped with, and a dual field $4$-form $G={}^\star\Phi$, in terms of which the metric can be constructed, which corresponds to the field strength of a gauge potential. Writing $G=G_0+d\Gamma$, then $\Gamma$ is a $3$-form under which the membrane is charged.\\

A non-perturbative formulation of the topological string naturally unifies branes and fields of the A and B models. In the $7$-dimensional context the unification is achieved recovering, near a boundary with normal direction $dt$, the $3$-form $\Phi$ and the $4$-form $G$, which define the $G_2$ structure, and combining the fields of the A and B models on the boundary according to:
\begin{equation}
\Phi= \mathfrak{Re} \, \Omega + k \wedge dt\,, \qquad \qquad G= \mathfrak{Im} \, \Omega \wedge dt + \frac{1}{2} k \wedge k\,.
\end{equation}


\subsection{CS gauge theory and 3D gravity}

We may inspect the paradigmatic case of the Einstein-Hilbert theory of gravity in three-dimensions, with cosmological constant, which we already introduced in the previous section. This is a topological theory without propagative degrees of freedom (no gravitons), which for Euclidean manifolds $\mathcal{M}_3$ is expressed by the action:
\begin{equation}
\mathcal{S}_{\rm 3-grav}=\int_{\mathcal{M}_3} \sqrt{g}\left( R- 2 \Lambda \right)\,.
\end{equation}
This is expressed, in the first order formalism, as:
\begin{equation}
\mathcal{S}_{\rm 3-grav}=\int_{\mathcal{M}_3}
{\rm Tr}\left( e \wedge F +\frac{\Lambda}{3} e\wedge e\wedge e \right)
\,,
\end{equation}
with $F=dA + A\wedge A$ being the field strength of the SU$(2)$ connection $A^i$ --- here $i=1,2,3$ is an internal index, in the adjoint representation  of the $\mathfrak{su}(2)$ Lie algebra --- and $e^i$ a $\mathfrak{su}(2)$-valued one-form on $\mathcal{M}_3$, such that $g_{ab}=(1/2){\rm Tr}(e_a e_b)$. The theory fulfils the equations of motion:
\begin{equation}
d_A e=0\,, \qquad  \qquad  F+ \Lambda e\wedge e\,,
\end{equation}
with $d_A$ the covariant derivative with respect to $A$, which provides, respectively, the metric compatibility condition and the Einstein equations with the cosmological constant. The Euclidean theory of gravity can be then reformulated in terms of a CS gauge theory as:
\begin{equation}
\mathcal{S}=\int_{\mathcal{M}_3} {\rm Tr}\left(
\mathcal{A} \wedge d\mathcal{A} + \frac{2}{3} \mathcal{A} \wedge \mathcal{A}\wedge \mathcal{A}
\right)\,,
\end{equation}
with $\mathcal{A}$ the gauge connection of a Lie group G, which is determined by the cosmological constant and can be regarded as the isometry group of the underlying geometric structure. For the Euclidean theory under scrutiny, depending on the sign of the cosmological constant, the gauge group can be: i) G=SL$(2,\mathbb{C})$, for $\Lambda <0$; G=ISO(3), for $\Lambda = 0$; G=SU$(2)\times$SU$(2)$, for $\Lambda >0$. For this theory, the Einstein equations ensure the flatness of the gauge connection $\mathcal{A}$, i.e. $d\mathcal{A}+ \mathcal{A} \wedge \mathcal{A}=0$.\\

The quantum version of three-dimensional gravity can be accomplished resorting to discretization methods. For instance, picking up a simplicial triangulation $\Delta \subset M$, to each tetrahedron is associated a quantum $6j$-symbol and hence the Turaev-Viro state sum $TV(\Delta)$, which reads:
\begin{equation}\label{TV}
TV(\Delta)= \left(- \frac{ (q^{1/2} - q^{-1/2} )^2}{2k}\right)^V \sum_{j_e} \prod_{\rm edges} [2j_e+1]_q \prod_{\rm tetrahedra} (6j)_q\,,
\end{equation}
with $V$ total number of vertices in the triangulation, and $[2j_e+1]_q$ the quantum dimension of the spin $j$ representation of SU$(2)_q$, defined by:
\begin{equation}
[n]_q=\frac{q^{n/2}- q^{-n/2}}{q^{1/2}- q^{-1/2}}
\,.
\end{equation}
The Turaev-Viro model is independent on the triangulation, namely $TV(\mathcal{M}_3)= TV(\Delta)$, and hence the partition function in Eq.~\eqref{TV} provides a topological invariant \cite{TV}. It was then shown in \cite{tu:91} that $TV(\mathcal{M}_3)$ equals the square of the partition function of SU$(2)$ CS theory, namely the Reshetikhin-Turaev-Witten invariant \cite{WittenJonesPolinomials} $\mathcal{Z}_{\rm SU(2)}(\mathcal{M}_3)$:
\begin{equation}
TV(\mathcal{M}_3)= |\mathcal{Z}_{\rm SU(2)}(\mathcal{M}_3)|^2\,.
\end{equation}

\subsection{Plebanski gravity in 4D}
An expression of the Einstein-Hilbert theory of gravity on a four dimensional manifold $\mathcal{M}_4$ was provided by Plebanski \cite{Pleb:75} within the framework of topological quantum field theory:
\begin{eqnarray}
\mathcal{S}_{\rm 4D}= \!\!\! &&
\int_{\mathcal{M}_4} \Sigma^{IJ} \wedge F_{IJ} - \frac{\Lambda}{24}  \Sigma^{IJ} \wedge  \Sigma_{IJ} + \Psi_{IJKL}   \Sigma^{IJ} \wedge  \Sigma^{KL}  \nonumber\\
=\!\!\! &&\int_{\mathcal{M}_4} \Sigma_+^k \wedge F^+_k - \frac{\Lambda}{24}  \Sigma_+^k \wedge  \Sigma^+_k + \Psi^+_{ij}   \Sigma_+^i \wedge  \Sigma_+^j  \nonumber\\
&+&\int_{\mathcal{M}_4} \Sigma_-^k \wedge F^-_k - \frac{\Lambda}{24}  \Sigma_-^k \wedge  \Sigma^-_k + \Psi^-_{ij}   \Sigma^i_- \wedge  \Sigma^j_- \,,
\end{eqnarray}
with $I,J=1,2,3,4$ indices of the fundamental representation of Spin$(4)$=SU(2)$\times$SU(2) and $A^{IJ}=(A_+^k, A_-^k)$ Spin$(4)$ gauge connection on the Euclidean space $\mathcal{M}_4$, with curvature:
\begin{equation}
F^{IJ}=(F^k_+, F^{k'}_-)\,, \qquad \qquad
F_\pm^k=dA_\pm^k + \epsilon^{i}_{\ jk} A_\pm^j A_\pm^k
\,,
\end{equation}
where $\Lambda$ is the cosmological constant and $\Psi^\pm_{ij}=\Psi^\pm_{(ij)}$ are symmetric SU(2)-tensors. The equations of motion elucidate the connection with the Einstein-Hilbert action. Varying with respect to $\Psi_{ij}$, the simplicity constraint is recovered:
\begin{equation}
\Sigma^{(i}\wedge \Sigma^{j)} -\frac{1}{3} \delta^{ij} \Sigma^k \wedge  \Sigma_k=0\,,
\end{equation}
which yields the solutions:
\begin{equation}
\Sigma^{IJ}=\frac{1}{2}\epsilon^{IJ}_{\ \ KL} e^K \wedge e^L + \frac{1}{2\gamma} \delta^{IJ}_{\ \ KL} e^K \wedge e^L
\,,
\end{equation}
with $\gamma$ a real parameter.

\subsection{K\"{a}hler and Kodaira-Spencer gravity in 6D}

In a similar way, in a six-dimensional manifold one can consider two different form theories of gravity, within a $\mathcal{N}=2$ Topological String Theory.  First, the K\"ahler theory of gravity \cite{6Da} provides a description of the target space gravity, namely String Field Theory, of the topological A model. Its action is defined by:
\begin{equation}
\mathcal{S}_{\rm Kahler}=\int_{\mathcal{M}_6} \left(
\frac{1}{2} K \frac{1}{ (\partial -\bar{\partial})^\dagger} dK +\frac{1}{3} K\wedge K \wedge K
\right)\,,
\end{equation}
with $K$ variation of the complexified K\"{a}hler form on $M$.  The action of K\"{a}hler gravity is invariant under the gauge transformation:
\begin{equation}
\delta_\alpha K = d\alpha - (\partial -\bar{\partial})^\dagger (K\wedge \alpha)\,,
\end{equation}
being $\alpha$ a one-form on $M$ such that $(\partial -\bar{\partial})^\dagger \alpha=0$.  One derives for K\"{a}hler gravity the equations of motion:
\begin{equation}
dK + (\partial -\bar{\partial})^\dagger (K\wedge K)=0\,,
\end{equation}
yielding, as we are going to observe also for the Kodaira-Spencer theory, the decomposition of $K$ into massless and massive modes:
\begin{equation}
K=x +  (\partial -\bar{\partial})^\dagger \gamma\,, \qquad \qquad x \in H^{1,1}(\mathcal{M}_6, \mathbb{C})\,,
\end{equation}
where $x \in H^{1,1}(\mathcal{M}_6, \mathbb{C})$ denotes the K\"{a}hler moduli, not integrated over, and $\gamma\in \Omega^3(\mathcal{M}_6)$ encodes the massive modes of $K$. The K\"{a}hler action can be then expressed, introducing the short notation $d^c=(\partial -\bar{\partial})$,  as:
\begin{equation}
\mathcal{S}_{\rm Kahler} = \int_{\mathcal{M}_6} \left(
\frac{1}{2} d\gamma \wedge d^{c \dagger} \gamma
+\frac{1}{3} K\wedge K \wedge K
\right)\,.
\end{equation}
D-branes of the A model are charged under $\gamma$, as it happens in the B model. Consequently, branes are sources for $K$, and modify its integral on the 2-cycles that interconnect them.\\

The Kodaira-Spencer theory of gravity \cite{6Db} is, on the other side, the string field theory of the topological B model. It provides a description of the variations of the complex structure. The basic field is a vector-valued one-form field $A$, with dynamics dictated by:
\begin{equation}
\mathcal{S}_{\rm KS}=\frac{1}{2}\int_{\mathcal{M}_6} A' \frac{1}{\partial} \bar{\partial}A' +\frac{1}{6} \int_{\mathcal{M}_6} (A\wedge A)' \wedge A' \,,
\end{equation}
where $A'=(A\cdot \Omega_0)$ denotes the product with the background holomorphic $(3,0)$ form. A variation of $\Omega$ is then defined by the formula in terms of $A$, namely:
\begin{equation}
\Omega=\Omega_0+A'+(A\wedge A)'+ (A\wedge A \wedge A)'\,,
\end{equation}
with $\Omega=\Omega_{ijk} dz^i \wedge dz^j \wedge dz^k$. Because of the non-local term in the action, $A$ must fulfil $\delta A'=0$, which provides the solution:
\begin{equation}
A'=x+\partial \phi\,, \qquad \qquad x\in H^{2,1}(\mathcal{M}_6, \mathbb{C})\,,
\end{equation}
$H^{2,1}(\mathcal{M}_6, \mathbb{C})$ representing the frozen (in the Kodaira-Spencer theory) massless modes (moduli) of $\Omega$, and $\phi\in \Omega^{1,1}(\mathcal{M}_6, \mathbb{C})$ the dynamical massive mode degrees of freedom. In terms of the $x$ and $\phi$ modes, the theory can be recast as:
\begin{equation}
\mathcal{S}_{\rm KS}=\frac{1}{2}\int_{\mathcal{M}_6} \partial \phi \wedge \bar{\partial} \phi +\frac{1}{6} \int_{\mathcal{M}_6} (A\wedge A)' \wedge A' \,,
\end{equation}
which yields the equations of motion:
\begin{equation}\label{eom}
\bar{\partial} A'+{\partial} (A \wedge A)'=0\,.
\end{equation}
The condition $\partial A'=0$, together with \eqref{eom}, implies that the holomorphic three-form is closed on-shell, namely:
\begin{equation}
d\Omega=0\,.
\end{equation}
Looking at $\phi$ as a dynamical degree of freedom, we realize the existence of a large shift symmetry,
\begin{equation}\label{lss}
\phi \rightarrow \phi + \epsilon\,,
\end{equation}
with $\partial \epsilon=0$, which can be used to gauge the anti-holomorphic part of $\phi$ to zero, namely $\bar{\partial} \phi=0$. The field $\phi$ is then the equivalent of chiral boson in two-dimension, making the Kodaira-Spencer theory chiral. One can also see that D1-branes within the B model are charged directly under $\phi$, as one can realize considering a D1-brane wrapped on a 2-cycle $E$ moving towards another 2-cycle $E'$, and noticing that there is a 3-chain $C$ that interpolates between $E$ and $E'$, and that the corresponding variation of the action reads:
\begin{equation}
\delta \mathcal{S}=\int_C \Omega = \int_C \partial \phi = \int_C d\phi = \int_E \phi - \int_{E'} \phi  \,,
\end{equation}
which introduces D1-branes as a source for $\Omega$, and hence induces the modified equations of motion:
\begin{equation}
\bar{\partial} A' = \bar{\partial} \partial  \phi =\delta^4_E \,,
\end{equation}
turning the kinetic term $\phi  \partial \bar{\partial} \phi $ into $\int_E \phi$.  The D1-branes coupling to $\phi$ implies that the amplitudes that involve D1-brane instantons are sensitive to the shift that enters the non-perturbative definition of the B model, thus that the partition function of the B model is a non-perturbative function of $ x \in H^{2,1}(\mathcal{M}_6)$ and $b\in H^{1,1}(\mathcal{M}_6,\mathbb{C})$.

\subsection{3-form and 4-form actions in 7D}

On $7$-dimensional manifolds $\mathcal{M}_7$, one can introduce a stable three-form $\Phi\in\Omega^3(\mathcal{M}_7, \mathbb{R})$, which determines a $G_2$ structure on $\mathcal{M}_7$ --- $G_2$ is a subgroup of GL$(7, \mathbb{R})$ that fixes at each point $\Phi$ --- and enables the expression for the metric $g$:
\begin{equation}
g_{ij}=B_{jk} {\rm det}(B)^{-1/9}\,,
\end{equation}
with:
\begin{equation}
B_{jk}=-\frac{1}{144} \Phi_{j i_1 i_2} \Phi_{k i_3 i_4} \Phi_{i_5 i_6 i_7} \epsilon^{i_1 i_2 i_3 i_4 i_5 i_6 i_7}\,.
\end{equation}
One can then write a functional that is the volume of $\mathcal{M}_7$ as determined by $g$, namely:
\begin{equation}
V_7(\Phi)=\int_{\mathcal{M}_7} \sqrt{g \Phi} = \int_{\mathcal{M}_7} {\rm det}(B)^{1/9}\,.
\end{equation}
The 3-form field $\Phi$ individuates the metric and any other derived quantities, including the Hodge operator $\star_\Phi$, which allows to express the potential as:
\begin{equation}
V_7(\Phi)=\int_{\mathcal{M}_7} \Phi \wedge \star_\Phi \Phi \,.
\end{equation}
The field $\Phi$ can be regarded as the field strength of a two-form, similarly as we have seen on the lower dimensional manifolds. Assuming it is closed, and varying it in a fixed cohomology class $[\Phi]\in H^3(\mathcal{M}_7, \mathbb{R})$, one finds a similar shift symmetry as in lower dimensional manifolds, i.e.
\begin{equation}
\Phi=\Phi_0 +dB\,,
\end{equation}
with $d\Phi_0=0$ and $B$ and arbitrary and real two-form on $\mathcal{M}_7$.  Critical points on $V_7(\Phi)$ on a fixed cohomology class provides the equations of motion that define closed and co-closed three-forms,
\begin{equation} \label{ppp}
d\Phi=0\,, \qquad \qquad d_{\star_\Phi}\Phi=0\,,
\end{equation}
fulfilling which $\Phi$ can be recognized to be an associative three-form for a $G_2$ holonomy metric on $\mathcal{M}_7$.

The $G_2$ holonomy condition can be also realized from a dual action involving a four-form $G$, namely the volume potential:
\begin{equation}
V_7(G)=\int_{\mathcal{M}_7} G\wedge_{\star_G} G\,.
\end{equation}
This specific realization is identified in \cite{Dijkgraaf:2004te} as the effective action of $7$-dimensional topological M-theory. Varying the four-form $G$ in a fixed cohomology class $[G]\in H^4(\mathcal{M}_7,\mathbb{R})$, one immediately recover that $G=G_0 +d\Gamma$, with $\Gamma$ an arbitrary real three-form on $\mathcal{M}_7$ and $G_0$ closed. The conditions for the extremization of the action $V_7(G)$ then write:
\begin{equation}
dG=0\,, \qquad \qquad d_{\star_G}G=0\,,
\end{equation}
which is another way with respect to Eq.~\eqref{ppp} to express that $\mathcal{M}_7$ has a $G_2$ holonomy, that can be then recast in terms of the co-associative four-form $G=\star_{\Phi} \Phi$.\\



TQFTs on base manifolds higher than three dimensions, which we reviewed in Sec.~\ref{qecc}, can be naturally accounted as an implementation of QECC within the framework of M-theory, which provides a general theoretical for this purpose. We leave to forthcoming studies detailed investigations on this subject.

\section{Conclusion}

In the accompanying Part I, we have shown here that multi-agent interactions meeting the requirements to implement a LOCC protocol generically induce QECCs. Here we have shown how spacetimes can, again generically, be regarded as QECCs. At this purpose, we have discussed in details the three-dimensional case, and then commented on possible extensions to higher dimension that involves notable constructions in extended (Plebanski) TQFT theories, and M-Theory. This yields an intriguing hypothesis, one consistent with remarks already made in \cite{addazi:21}: that the fundamental role of spacetime in physics is to provide ``a place to put'' redundancy.  Indeed it suggests that the concept of redundancy as a necessary resource for communication and the concept of spacetime as a necessary resource for dynamics may, at some fundamental level, be the same concept.  This being the case, it would substantially support Wheeler's contention \cite{wheeler:83} that physics is fundamentally about information exchange; a striking assertion that the methods of this paper point towards with significance. The overall train of ideas further suggests support for the pronouncement of Grinbaum \cite{grinbaum:17} that physics is fundamentally about language.


\section*{Acknowledgments}
A.M.~wishes to acknowledge support by the Shanghai Municipality, through the grant No.~KBH1512299, by Fudan University, through the grant No.~JJH1512105, the Natural Science Foundation of China, through the grant No.~11875113, and by the Department of Physics at Fudan University, through the grant No.~IDH1512092/001.

\end{document}